\documentclass{aa}  
\usepackage{graphicx}
\usepackage{lscape}
\usepackage{natbib} 
\usepackage{siunitx}
\usepackage{url}
\usepackage[varg]{txfonts}
\usepackage{siunitx}

\def\changed{}

\bibpunct{(}{)}{;}{a}{}{,} 
\defcitealias{HK2000}{HK2000}
\defcitealias{Hamann2006}{HGL06}
\defcitealias{BAT99}{BAT99}
\begin{document}

\title{Stellar population of the superbubble N\,206 in the LMC}

   \subtitle{I. Analysis of the Of-type stars}

   \author{Varsha Ramachandran\inst{1}
          \and R. Hainich \inst{1}
          \and W.-R. Hamann\inst{1}
          \and L. M. Oskinova\inst{1}
          \and T. Shenar\inst{1}
          \and A. A. C. Sander\inst{1}
          \and H. Todt\inst{1}
          \and J. S. Gallagher\inst{2} 
          }

   \institute{Institut f\"ur Physik und Astronomie,
              Universit\"at Potsdam,
              Karl-Liebknecht-Str. 24/25, D-14476 Potsdam, Germany \\
              \email{varsha@astro.physik.uni-potsdam.de}
              \and Department of Astronomy, University of Wisconsin - Madison, 
WI, USA }       
   \date{Received <date> / Accepted <date>}

\abstract
%
%
{ Massive stars severely influence their environment by their strong 
ionizing radiation and by the momentum and kinetic energy input provided by 
their stellar winds and supernovae. Quantitative analyses of massive 
stars are required to understand how their feedback creates and shapes 
large scale structures of the interstellar medium. The giant H\,{\sc ii} region N\,206 in the 
Large Magellanic Cloud contains an OB association that powers
a superbubble filled with hot X-ray emitting gas, serving as an ideal laboratory in this context.
} 
%
%
{We aim to estimate stellar and wind parameters of all OB stars in N\,206 by means of quantitative spectroscopic analyses. In this first paper, we focus on the nine Of-type  stars located in this region. We determine their ionizing flux and wind mechanical energy. The analysis of nitrogen abundances in our 
sample probes rotational mixing.  
}
%
%
{
We obtained optical spectra with the multi-object spectrograph FLAMES at the 
ESO-VLT. When possible, the 
optical spectroscopy was complemented by UV spectra from the HST, IUE, 
and FUSE archives. Detailed spectral classifications are presented for our sample 
Of-type stars. For the quantitative spectroscopic analysis we used
the Potsdam Wolf-Rayet (PoWR) model atmosphere code. We determined the physical parameters 
and nitrogen abundances of our sample stars by fitting synthetic 
spectra to the observations.
} 
%
%
{
 \changed{The stellar and wind parameters of nine Of-type stars, which are largely  derived from spectral 
analysis are used to construct wind momentum\,$-$\,luminosity relationship. We find that our sample follows a relation close to the theoretical prediction, assuming clumped winds.}
The most massive star in the N\,206 association is an Of supergiant that has a 
very 
high mass-loss rate. Two objects in our sample reveal composite spectra, showing 
that the Of primaries have companions of late O subtype. All stars in our sample 
have an evolutionary age of less than 
4 million years, with the O2-type star being the youngest.
All these stars show a systematic discrepancy between evolutionary and 
spectroscopic masses. All stars in our 
sample are nitrogen enriched. Nitrogen enrichment shows a clear correlation with increasing projected rotational velocities. 
}
%
%
{
 The mechanical energy input from the Of stars alone is comparable to the 
energy stored in the N\,206 superbubble as measured from the observed X-ray and 
H$\alpha$ 
emission.
}


\maketitle

\section{Introduction}
\label{sect:intro}

Superbubbles filled by  hot, $\sim 1$\,MK, gas are the large scale 
structures with characteristic size $\sim 100$\,pc in the interstellar medium 
(ISM) \citep{Maclow1988}. Superbubbles provide direct evidence 
for the energy feedback from massive star clusters or OB associations to the 
ISM.  With a distance 
modulus of only $D\!M$\,=\,18.5\,mag \citep{Madore1998,Pietrzynski2013}, the 
Large Magellanic Cloud (LMC) provides a good platform for detailed spectroscopy 
of massive stars in superbubbles. Additionally, its face-on aspect and low 
interstellar extinction make it an excellent laboratory to study feedback. With 
an observed metallicity of approximately [Fe/H]\,=\,$-0.31 \pm 0.04$ or 0.5 
$Z_{\odot}$ \citep{Rolleston2002}, the LMC exhibits a very different history of 
star formation compared to the Milky Way. Several massive star-forming regions 
in the LMC, especially 30 Doradus and the associated stellar populations have 
been extensively studied by many authors 
\citep[e.g.,][]{Evans2011,Ramirez-Agudelo2017}. In this paper we present 
the study of the Of-type stars in the massive star-forming region N\,206.

N\,206 (alias LHA\,120-N\,206 or DEM\,L\,221) is a giant \ion{H}{ii} region in 
the south-east of the LMC that is energized by the young cluster NGC\,2018 
(see Fig.\,1) and two OB associations, LH\,66 and LH\,69. Imaging studies with 
{\em HST}, {\em Spitzer}, and {\em WISE} in the optical and infrared (IR) have 
already unveiled the spatial structure of the NGC\,2018/N\,206 region 
\citep{Gorjian2004,Romita2010}. This complex contains a superbubble observed 
as a source of diffuse X-rays, and a supernova remnant. The current star 
formation in N\,206 is taking place at the rim of the X-ray superbubble 
\citep{Gruendl2009}. The X-ray emission in N\,206 region has been studied 
by \citet{Kavanagh2012} using observations obtained by the \emph{XMM-Newton} 
X-ray telescope. However, the study of feedback was constrained by the 
lack of detailed knowledge on its massive star population. 

To obtain a census of young massive stars in N\,206 complex and study their 
feedback we conducted a spectroscopic survey of all blue stars with $m_{V} < 
16$\,mag  with the FLAMES multi-object spectrograph at ESO's Very 
Large Telescope (VLT). Our total sample comprises 164 OB-type stars. The inspection 
and classification of the spectra revealed that nine objects belong to the Of 
subclass \citep{Walborn2002} and comprise the hottest stars in our whole sample. 
The present paper focuses on the spectral analysis of these Of-type stars, while 
the subsequent paper (Paper\,II) will cover the detailed analyses of the entire 
OB star population, along with a detailed investigation of the energy 
feedback in this region.

The spectroscopic observations and spectral classifications are presented in 
Sect.\,\ref{sect:spec}. Section\,\ref{sect:analysis} describes the quantitative 
analyses of these stellar spectra using Potsdam Wolf-Rayet (PoWR) atmosphere 
models. Section\,\ref{sect:results} presents the results and discussions. The 
final Sect.\,\ref{sect:summary} provides a summary and general conclusions. The 
appendices encompass comments on the individual objects (Appendix A) and the 
spectral fits of the analyzed Of stars (Appendix B).
\section{Spectroscopy}
\label{sect:spec}

We observed the complete massive star population associated with the N\,206 
superbubble on 2015 December 19-20 with VLT-FLAMES. In the Medusa-fiber mode, 
FLAMES \citep{Pasquini2002} can simultaneously record the spectra of up to 132 
targets. Each fiber has an aperture of $1.2\arcsec$ radius. The nine Of 
stars are a subsample of this larger survey. The observation was carried out 
using three of the standard settings of the Giraffe spectrograph LR02 (resolving 
power $R$\,=\,6000, 3960--4567\,\AA), LR03 ($R$\,=\,7500, 4501--5071\,\AA), and 
HR15N ($R$\,=\,19200, 6442--6817\,\AA), respectively. We took three to six 
exposures (30 minutes each) for each pointing in three spectrograph setting 
to improve the S/N. The higher resolution at H$ \alpha $ is utilized to 
determine stellar wind parameters and to distinguish nebular emission from the 
stellar lines.

The ESO Common Pipeline Library\footnote{http://www.eso.org/observing/cpl} 
FLAMES reduction routines  were executed for the standard processing stages such 
as bias subtraction, flat fielding, and wavelength calibration. We used the ESO 
data file organizer GASGANO\footnote{VLT-PRO-ESO-19000-1930/1.0/27-Sep-99 VLT 
Data Flow System, Gasgano DFS File Organizer Design Document} for organizing and 
inspecting the VLT data and to execute the data reduction tasks. The obtained 
spectra are not flux calibrated. Multi-exposures were normalized and median 
combined to get the final spectra without cosmic rays in each settings. \changed{The spectra were rectified by fitting the stellar continuum with a piece-wise linear function.} Finally, 
for each star, the LR02 and LR03 spectra were merged to form the medium 
resolution blue spectra from 3960 to 5071\,\AA. The sky background was 
negligibly small compared to the stellar spectra. The spectral data reduction 
was carried out without nebular background subtraction. Therefore, stars with a bright 
background might show nebular emission lines such as H$\alpha$, [\ion{O}{iii}], 
[\ion{N}{ii}] and [\ion{S}{ii}] superimposed on the stellar spectra.

\begin{figure*}
 \centering
 \includegraphics[scale=0.5,trim={1cm 0 0 0}]{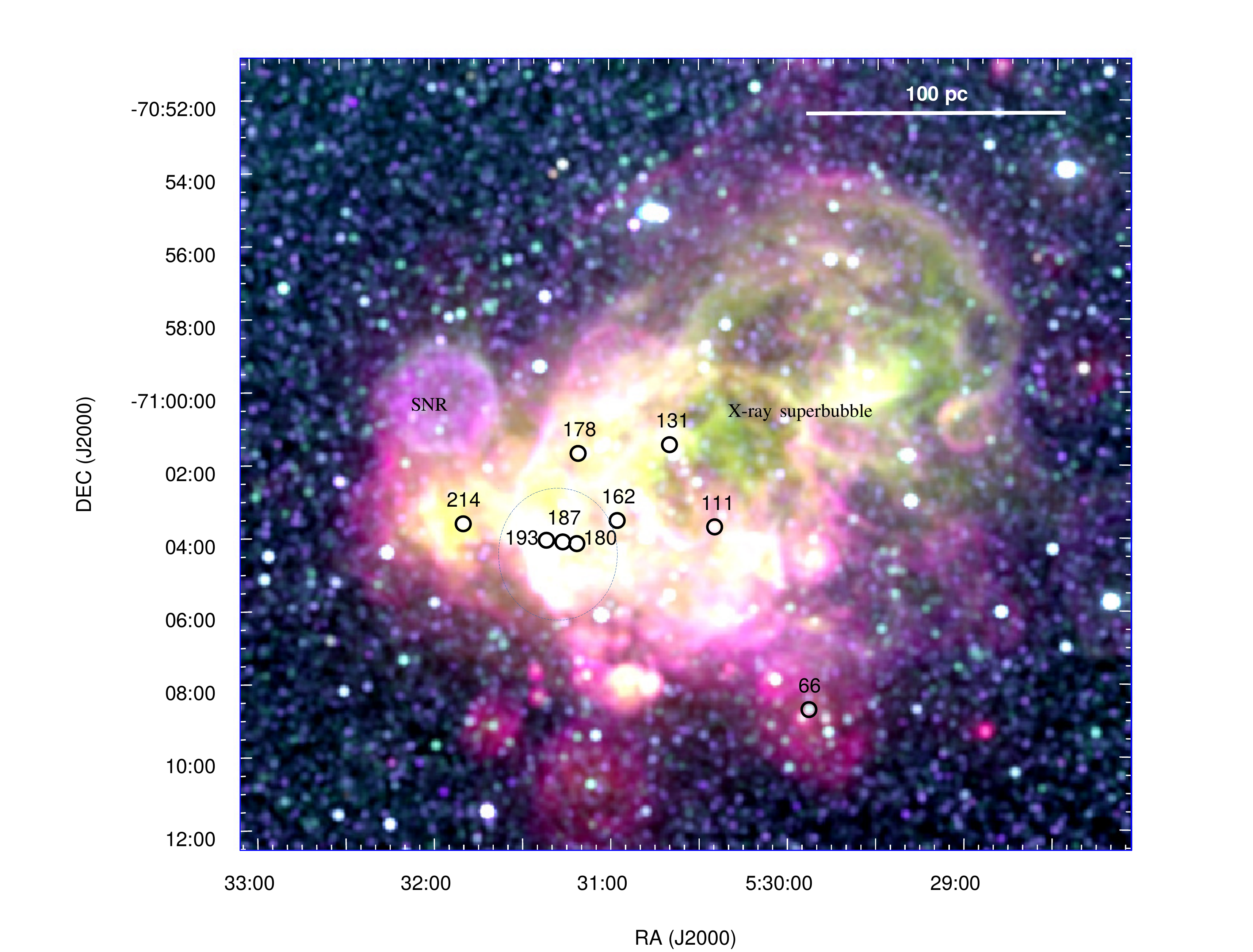}
 \caption{Location of the Of-type stars in the N\,206 \ion{H}{ii} region. Three 
color composite image (H$\alpha$ (red) + [\ion{O}{iii}] (green) + [\ion{S}{ii}] 
(blue)) shown in the background is from the Magellanic Cloud Emission-Line 
Survey \citep[MCELS; ][]{Smith2005}. The Of stars studied here are marked with 
the N206-FS number corresponding to Table\,\ref{table:ofstars}. The dotted 
circle (blue) shows the rough locus of young cluster NGC\,2018. The superbubble 
is located north-west and the SNR is at north-east of NGC\,2018. } 
 \label{fig:Halpha_n206}
 \end{figure*}

We obtained a total of 234 spectra with good signal-to-noise (S/N$>$\,50).
The whole sample encompasses the spectra of 164 OB stars. Additional fibers were 
placed on the H\,{\sc ii} region, X-ray bubble, and supernova remnant. We assigned a 
naming convention for all the objects with N206-FS (N\,206 FLAMES Survey) and a 
number corresponding to ascending order of their right-ascension (1--234). 
Paper\,II (Ramachandran et al. in prep.) will publish the catalog in detail. We 
identified and analyzed nine Of-type stars among this sample. Their positions are 
marked on the color composite image of N\,206 in Fig.\,\ref{fig:Halpha_n206}.

Ultraviolet (UV) spectra are available for three Of stars in our sample, and we retrieved these 
from the Mikulski Archive for Space Telescopes (MAST\footnote{ 
http://archive.stsci.edu/}). For N206-FS\,187, a HST/Space Telescope 
Imaging Spectrograph (STIS) spectrum (ID: O63521010) exists. This was taken with the E140M 
grating (aperture $ 0.2\arcsec \times 0.2\arcsec$), covering the wavelength 
interval 1150--1700\,\AA, with an effective resolving power of $R$\,=\,46000. An 
International Ultraviolet Explorer (IUE) short-wavelength spectrum taken in high 
dispersion mode is available for N206-FS\,180 (ID: SWP14022). This spectrum was 
taken with a large aperture ($21\arcsec \times 9\arcsec$) in the wavelength 
range 1150--2000\,\AA. A far-UV Far Ultraviolet Spectroscopic Explorer (FUSE)
spectrum is available for N206-FS\,214 (ID: d0980601) in the wavelength range 
905--1187\,\AA, taken with a medium aperture ($ 4\arcsec \times 20\arcsec$).

In addition to the spectra, we used various photometric data from the VizieR 
archive to construct the spectral energy distribution (SED). Ultraviolet and optical ($U, 
B, V, $ and $ I$) photometry were taken from \citet{Zaritsky2004}. The infrared 
magnitudes ($JHK_{s}$ and \textit{Spitzer}-$IRAC$) of the sources are based on the 
catalog by \citet{Bonanos2009}.

\subsection*{Spectral classification}
\label{subsect:class}

The spectral classification of the stars is primarily based on the spectral 
lines in the range 3960--5071\,\AA. We mainly followed the classification 
schemes proposed in \citet{Sota2011}, \citet{Sota2014}, and \citet{Walborn2014}.

The main criterion for spectral classification is the \ion{He}{i}/\ion{He}{ii} 
ionization equilibrium. The diagnostic lines used for this purpose are 
\ion{He}{i} lines at 4471\,\AA, 4713\,\AA, and 4387\,\AA\, compared to the 
\ion{He}{ii} lines at 4200\,\AA\, and 4541\,\AA. However, this criterion leaves 
uncertainties in the subtype of the earliest O stars because of their weak or 
negligible \ion{He}{i} lines.

Moreover, the strength and morphology of the optical nitrogen lines 
\ion{N}{iii\,$\lambda\lambda$4634--4640--4642} (hereafter \ion{N}{iii}), 
\ion{N}{iv\,$\lambda4058$} (hereafter \ion{N}{iv}) and 
\ion{N}{v\,$\lambda\lambda$4604--4620} (hereafter \ion{N}{v}) are the 
fundamental classification criteria of the different Of spectral subtypes 
\citep{Walborn1971}. As suggested by \citet{Walborn2002}, we used the 
\ion{N}{iv\,$\lambda4058$}/\ion{N}{iii\,$\lambda4640$} (here after 
\ion{N}{iv}/\ion{N}{iii}) and \ion{N}{iv\,$\lambda4058$}/\ion{N}{v\,$\lambda 
4620$} (hereafter \ion{N}{iv}/\ion{N}{v}) emission line ratio as the primary   
criterion for the Of subtypes, instead of \ion{He}{i}/\ion{He}{ii}. The 
luminosity-class criteria of early O stars are mainly based on the strength of 
\ion{He}{ii}\,4686 and \ion{N}{iii} lines. 

\begin{figure*}

 \begin{center}
 \includegraphics[scale=0.7 ,angle=0,trim={2cm 1cm 6cm 0cm}]{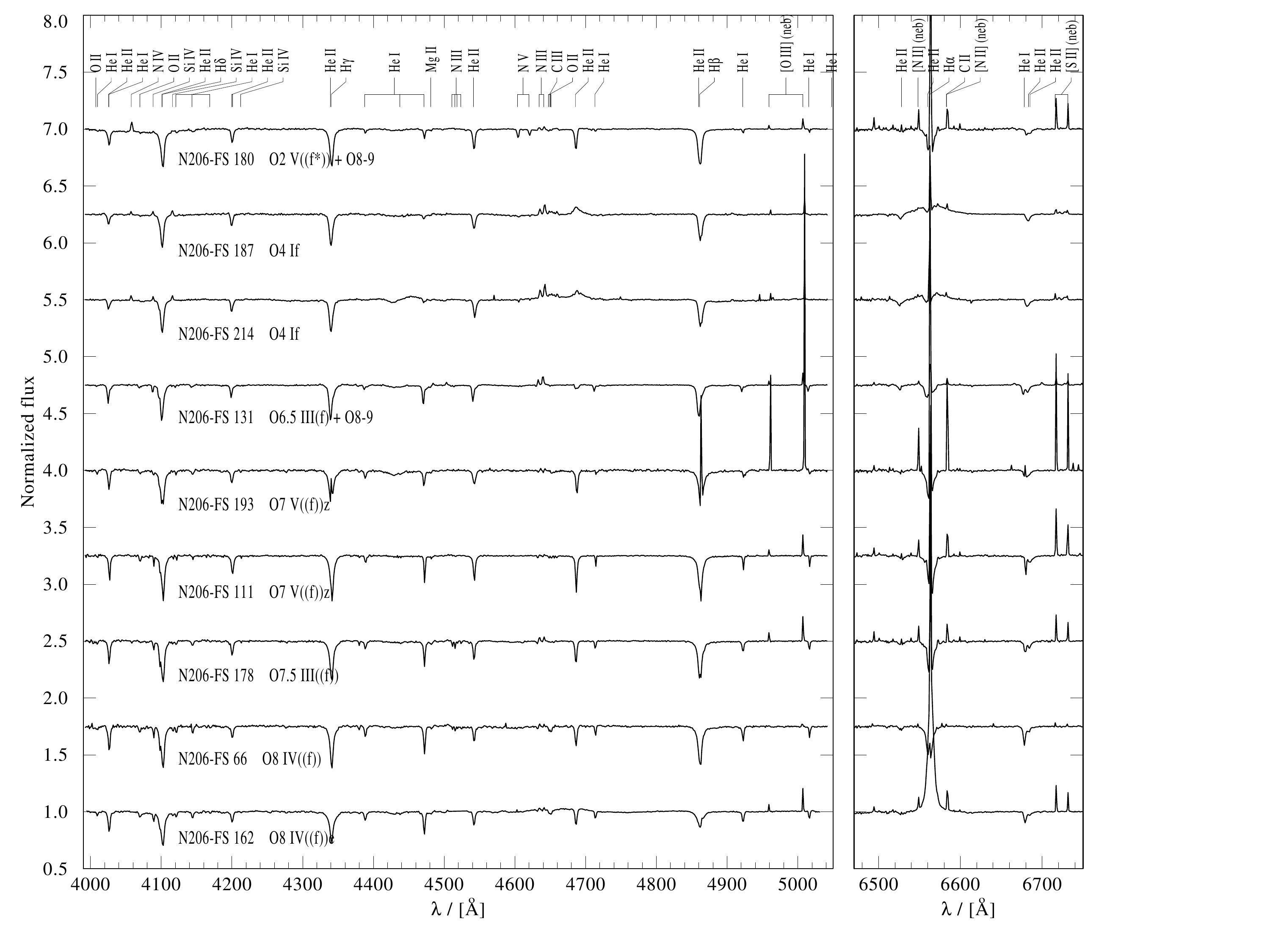}
 \end{center}
 \caption{Normalized spectra of the nine Of-type stars. The left panel depicts 
the medium resolution spectra in the blue (setting: LR02 and LR03). The high 
resolution spectra in the red (setting: HR15N) are shown in the right panel 
which includes H$\alpha$ line. The central emission in H$\alpha$ and the [\ion{O}{iii}], [\ion{N}{ii}], and [\ion{S}{ii}] lines are from nebular emission 
(neb). } 
 \label{fig:of_spectra}
 \end{figure*}

The spectral type and coordinates of the nine Of-type stars are given in 
Table\,\ref{table:ofstars}, together with their most prominent aliases. The 
corresponding normalized spectra are shown in Fig.\,\ref{fig:of_spectra}. In the 
following, we comment on the individual Of stars.

Our sample contains one star of very early spectral subtype O2 
(N206-FS\,180), which was classified as an O5 V by \citet{Kavanagh2012}. 
It is located in the bright cluster NGC\,2018. The characteristic features 
of the O2 class in the blue spectra are strong \ion{N}{v} absorption lines and the 
\ion{N}{iv} emission significantly stronger than the \ion{N}{iii} emission. This 
subtype was introduced by \citet{Walborn2002}, also discussed in 
\citet{Gonzalez2012b} and \citet{Walborn2014}. 
 The presence of strong \ion{He}{ii}\,4686  absorption shows that N206-FS\,180 
is a young main sequence star. According to the definition by 
\cite{Walborn2002}, the primary characteristic of the spectral type O2 V((f*)) 
in comparison to O3 V((f*)) is weak or absent \ion{N}{iii} emission. The 
spectrum of N206-FS\,180 clearly complies with this. The presence of \ion{He}{i} 
absorption lines in this very early O2 spectrum indicates a secondary component 
(see Sect.\,\ref{subsubsec:composite}).

\begin{figure}
 \centering
 \includegraphics[scale=0.53,trim={0 20cm 4cm 2cm}]{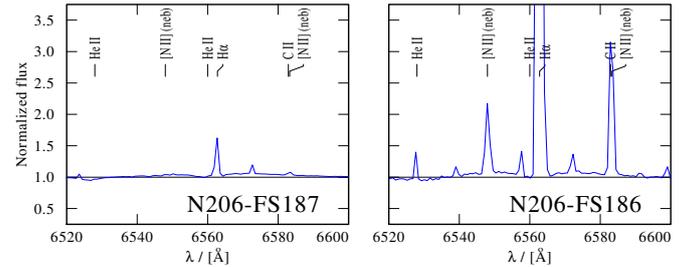}
 \caption{Comparison of H$\alpha$ lines of N206-FS\,187 (O4 If) and neighboring star N206-FS\,186 (O8.5 (V)e). } 
 \label{fig:186}
 \end{figure}

Two Of-type supergiants are present in the sample, N206-FS\,187 and 214. N206-FS\,187 was identified as binary/ multiple system (Of + emission line star) in \citet{Hutchings1980}. \citet{Hutchings1982} denoted this object as an emission 
line star. \citet{Kavanagh2012} classified this star as an O4-5 giant. We 
reclassified this spectrum as O4 If (see Table 3 and Section 3.1.3 of \cite{Sota2011} for more details of this classification scheme). This star is situated in the crowded region of the young cluster NGC 2018. We took the spectrum of N206-FS\,186 (O8.5 (V)e), that is $ \approx $ 2.5\arcsec\,away from N206-FS\,187. This emission line (late Oe type) star could have been wrongly identified as a companion in previous papers (see their H$\alpha$ lines in Fig.\,\ref{fig:186}).  Given the spacial proximity of the two stars, they were erroneously considered as a single source in several past studies.

 In both supergiants N206-FS\,187 and 214, the \ion{N}{iv} emission is much weaker than the \ion{N}{iii} emission, and the \ion{N}{v} absorption is negligible, which indicates an O4 spectral type. Also, the \ion{He}{i} absorption lines are almost absent or negligible in the spectra. The significant \ion{He}{ii}\,4686 emission confirms its supergiant nature. In addition to this, narrow  \ion{Si}{iv\,$\lambda$4089--4116} emission features are also present in both spectra. 

Compared to this, N206-FS\,131 has no or very weak \ion{He}{ii}\,4686 absorption 
and is therefore classified as bright giant (luminosity class II). The spectrum of this 
star shows a strong \ion{N}{iii} emission line, but with no \ion{N}{v} and 
\ion{Si}{iv} emission lines, implying this star is cooler than previous stars 
(O6.5). This star is also suspected as a binary from the strength of the \ion{He}{i} lines. 
Another giant with strong \ion{N}{iii} emission is N206-FS\,178 but with 
comparatively strong \ion{He}{ii}\,4686 absorption (luminosity class III).

\begin{table*}
\caption{\changed{Spectral type and coordinates of the nine Of-type stars in the N\,206 
superbubble}} 
\label{table:ofstars}      
\centering

\begin{tabular}{cclll}
\hline
\hline
\noalign{\vspace{1mm}}
 N206-FS\tablefootmark{(2)} &RA (J2000)   & DEC (J2000)  & Alias names\tablefootmark{(1)} & Spectral type 
 \\
\# &(h:m:s)& (\degr : \arcmin : \arcsec)& &   \\
\noalign{\vspace{1mm}}
\hline 
\noalign{\vspace{1mm}}
66& 5:29:52.750 & $-$71:08:44.70  &	2MASS J05295273-7108446,  &   O8 IV((f)) 
  \\
 & & & [MLD95] LMC 1-263\\
111& 5:30:24.680 & $-$71:03:44.50  &	2MASS J05302469-7103445,  &   O7 V((f))z 
 \\ 
 & & & [MLD95] LMC 1-584\\
131 &5:30:39.870 & $-$71:01:29.30  &	HDE 269656,                  & 	O6.5 
II(f) + O8-9 \\
 & & &Sk $-$71 39   \\  
 162 &5:30:57.560 & $-$71:03:33.70  &	BI 189, [L63] 275,       & 	O8 
IV((f))e  \\
 & & & [MLD95] LMC 1-598\\ 
178&  5:31:10.670 & $-$71:01:43.30  &	2MASS J05311070-7101433, &	O7.5 
III((f))\\
 & & & [MLD95] LMC 1-716\\
180&  5:31:11.780 & $-$71:04:10.10  &	2MASS J05311185-7104101,  & 	O2 
V((f*)) + O8-9  \\
 & & & [MLD95] LMC 1-550\\
 187 & 5:31:15.650 & $-$71:04:10.00  &	HDE 269676,                 &	O4 If
   \\
 & & & Sk $-$71 45    \\  
  193& 5:31:21.470 & $-$71:04:05.80  &	2MASS J05312153-7104057,& 	O7 
V((f))z   \\
 & & & [MLD95] LMC 1-556 \\
  214  &  5:31:49.510 & $-$71:03:38.0   &	Sk $-$71 46,       & 	O4 If \\
 & & & 2MASS J05314960-7103381\\

\hline 
\end{tabular}
\tablefoot{\tablefoottext{1}{Taken from http://simbad.u-strasbg.fr/simbad/}
\tablefoottext{2}{The number refers to the whole catalog of objects in the 
N\,206 superbubble (Ramachandran et al. in prep.) based out of FLAMES survey 
(N206-FS). This contains 234 fiber positions in ascending order of their RA (N206-FS 
1--234). }}

\end{table*}

We discovered two stars that exhibit special characteristics typical for the Vz 
class, namely, N206-FS\,193 and N206-FS\,111.
According to \citet{Walborn2006}, these objects may be near or on the zero-age 
main sequence (ZAMS). As described in \citet{Walborn2014}, the main 
characteristic of the Vz class is the prominent \ion{He}{ii}\,4686 absorption 
feature that is stronger than any other \ion{He}{} line in the blue-violet region. 
Strong absorption in \ion{He}{ii}\,4686  corresponds to lower luminosity and a 
very young age (i.e., the inverse Of effect). These spectral features can be easily contaminated with strong nebular contribution, however.

Another interesting spectrum in the sample belongs to an Oef star, N206-FS\,162. 
The H$\alpha$ line is in broad emission while other Balmer lines are partially 
filled with disk emission,  which is a feature of a classical Oe/Be star. A very 
small \ion{N}{iii} emission indicates the `Of nature'. The star N206-FS\,66 is 
classified as a subgiant. It shows weak \ion{N}{iii} emission lines and a 
significant \ion{He}{ii}\,4686 absorption line. 

\section{The analysis}
\label{sect:analysis}

Our immediate objective is to determine the physical parameters of the 
individual stars. In order to analyze the FLAMES spectra, we calculated 
synthetic spectra using the Potsdam Wolf-Rayet (PoWR) model atmosphere code, and 
then fitted those to the observed spectra.

\subsection{The models}
\label{subsect:models}

The PoWR code solves the radiative transfer equation for a spherically expanding 
atmosphere and the statistical equilibrium equations simultaneously, while 
accounting for energy conservation and allowing for deviations from local 
thermodynamic equilibrium (i.e., non-LTE). Stellar parameters were determined 
iteratively. Since the hydrostatic and wind regimes of the atmosphere are solved 
consistently in the PoWR model \citep{Sander2015}, this code can be used for the 
spectroscopic analysis of any type of hot stars with winds, across a broad range 
of metallicities \citep{Hainich2014,Hainich2015,Oskinova2011,Shenar2015}. More 
details of the PoWR code are described in \citet{Graefener2002} and 
\citet{Hamann2004}.

A PoWR model is specified by its luminosity $L$, the stellar temperature 
$T_\ast$, the surface gravity $g_\ast$, and the
mass-loss rate $ \dot{M} $ as main parameters. PoWR defines the ``stellar 
temperature'' $T_\ast$ as the effective temperature referring to the stellar 
radius $R_\ast$ where, again by definition, the Rosseland optical depth reaches 
20. In the standard definition, the ``effective temperature'' $T_\mathrm{eff}$ 
refers to the radius $R_{2/3}$ where the Rosseland optical depth is 2/3,

\begin{equation}
\label{eq:sblaw}
L = 4 \pi \sigma_{\mathrm{SB}}\, R_\ast^2\, T_\ast^4~ \, = \, 4 \pi 
\sigma_{\mathrm{SB}}\, R_{2/3}^2\ \,T_{\mathrm{eff}}^4
\end{equation}

In the case of our program stars, the winds are optically thin and the differences between 
$T_\ast$ and $T_\mathrm{eff}$ are negligible. Since model spectra are most sensitive to  $T_*$, log\,$g_\ast$, $\dot M$, and $L$ these parameters are varied to find the best-fit model systematically.

In the non-LTE iteration, the line opacity and emissivity  profiles  are  
Gaussian  with  a  constant Doppler width $\varv_{\mathrm{Dop}}$. This 
parameter is set to 30\,km\,s$ ^{-1} $ for our `Of' sample. For the energy spectra, the Doppler velocity is split into the depth-dependent thermal 
velocity and a ``microturbulence velocity'' $\xi(r)$. We adopt  $\xi(r) = \rm 
max(\xi_{min},\, 0.1\varv(\emph{r}))$ for O-star models, where $\rm \xi_{min}= 20 $\,km\,s$ ^{-1} $ \citep{Shenar2016}.

Optically thin inhomogeneities in the model iterations are prescribed by the 
``clumping factor''  $D$ by which the density in the clumps is enhanced compared 
 to a homogeneous wind of the same $\dot{M}$ \citep{HK98}.
In the current study, we account for depth-dependent clumping assuming that the 
clumping starts at the sonic point, increases outward, and reaches a density 
contrast of $D=10$ at a radius of $R_{\mathrm{D}} = \, 10\,R_\ast$. 
Note that the empirical mass-loss rates when derived from H$\alpha$ emission 
scale with $D^{-1/2}$, since this line is mainly fed via recombination. We also 
varied the values of $D$ and $R_{\mathrm{D}}$, when necessary. Higher $D$ values 
and lower $R_{\mathrm{D}}$ values lead to a decrease in the mass-loss rates 
derived from H$\alpha$.

 The detailed form of the velocity field in the wind domain can affect 
spectral features originating in the wind. 
In the subsonic region, the velocity field is defined such that a hydrostatic 
density stratification is approached \citep{Sander2015}. In the supersonic 
region, the pre-specified wind velocity field $\varv(r)$ is assumed to follow 
the so-called $\beta$ -law \citep{CAK1975} 
\begin{equation}
\varv(r) = \varv_\infty \left( 1- \frac{r_{0}}{r} \right)^{\beta}.
\end{equation}
In this work, we adopt $\beta$=0.8, which is a typical value for O-type stars 
\citep{Kudritzki1989}.

The models are calculated using complex model atoms for H, He, C, N, O, Si, Mg, 
S, and P. 
The iron group elements (e.g., Fe, Ni) are treated with the so-called ``superlevel approach'' 
as described in \citet{Graefener2002}.

\subsection{Spectral fitting}
\label{subsec:specfit}

 The  spectral analysis is based on systematic fitting of observed spectra with 
grids of stellar atmosphere models. We constructed OB-star grids for LMC 
metallicity with the stellar temperature $T_\ast$ and the surface gravity 
log\,$g_\ast$ as parameters varied in the grid. Additional parameters such as
stellar mass $M$ and luminosity $L$ in the grid models are chosen according to 
the evolutionary tracks calculated by \citet{Brott2011}. 
The other parameters, namely the chemical composition and terminal wind 
velocity, are kept constant within one model grid. We also calculated some models 
with adjusted C, N, O, and Si abundance, when necessary. The LMC OB star grid\footnote{www.astro.physik.uni-potsdam.de/PoWR.html}
spans from $T_\ast$\,=\,13\,kK to 54\,kK with a spacing of 1\,kK, and log 
$g_\ast$\,=\,2.2 to 4.4 with a spacing of 0.2 dex.

 We proceeded as follows to derive the stellar and wind parameters. The main 
method to identify the stellar temperature is to fit the 
\ion{He}{i}/\ion{He}{ii} line ratio. For the present Of-type sample we also made 
use of \ion{N}{iii}/ \ion{N}{iv}/ \ion{N}{v} line ratios. The uncertainty in 
temperature determination according to the grid resolution is $ \pm $1\, kK. 
Since for hotter stars, the temperature determination mainly depends on the 
nitrogen ionization equilibrium, and the \ion{He}{i} lines are very weak, the uncertainty becomes larger ($ \pm 
$2-3\,kK). The surface gravity log\,$g_\ast$ was mainly determined from the 
wings of the Balmer lines, which are broadened by the Stark effect. Since the 
H$\alpha$ line is often affected by wind emission, we mainly used H$\gamma$ and 
H$\delta$ for this purpose. The typical uncertainty for log\,$g_\ast$ is 
$\pm$0.2 dex. The uncertainty in log\,$g_\ast$  also propagates to the 
temperature and gives a total uncertainty of $ \sim \pm $2\,kK.

\begin{figure*} 
\centering
\includegraphics[scale=0.6,trim={0 0cm 0cm 0}]{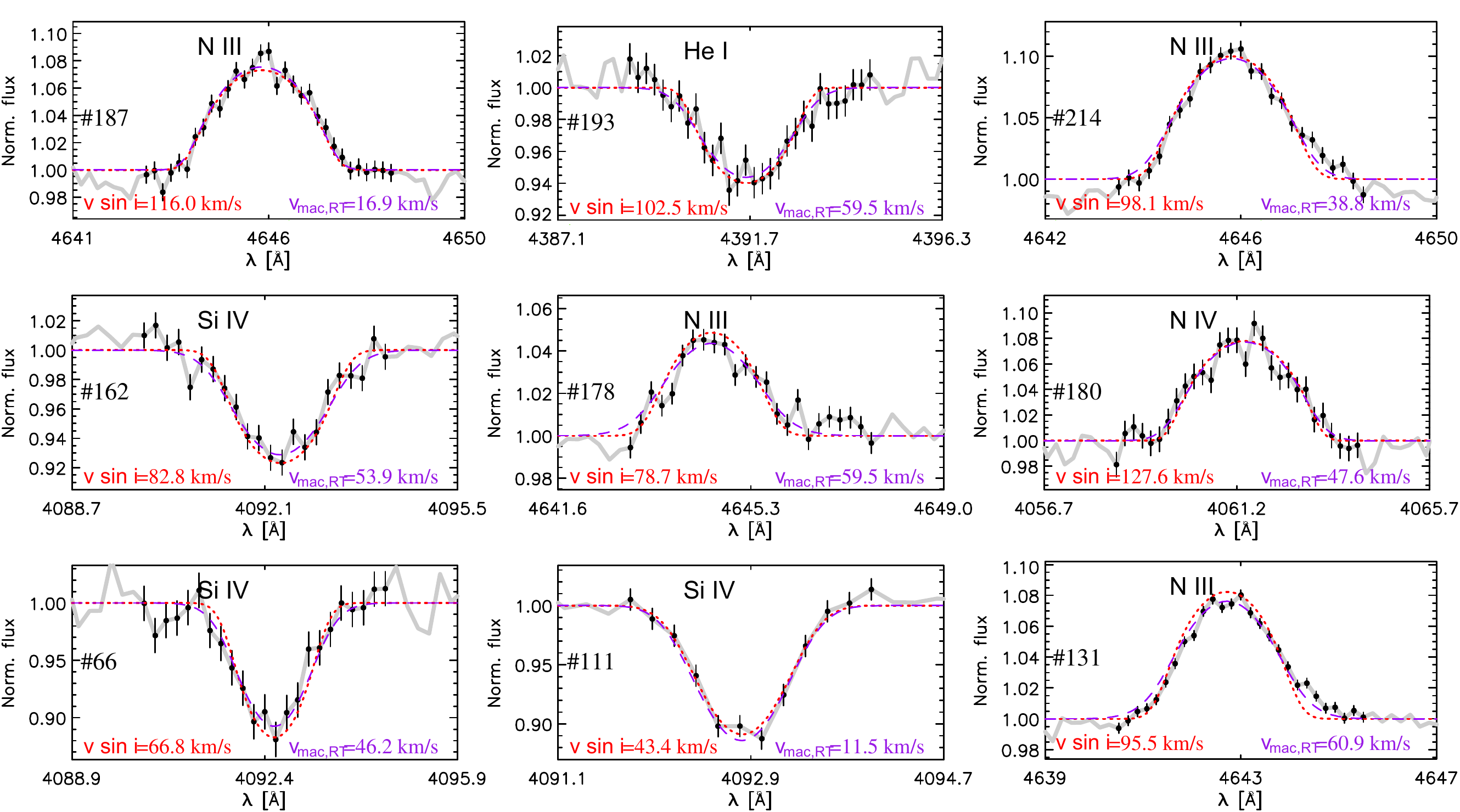}
\caption{Rotation velocity ($\varv$\,sin\,$i$) from the line profile fitting 
of Of-type stars using the \texttt{iacob-broad} tool. The gray curves show the 
observed line profile. The $\varv$\,sin\,$i$ is calculated from line profile 
fitting based on Fourier transform method (red dotted line) and goodness-of-fit 
analysis (violet dashed line). See text for more details.}
\label{fig:rotaionprof}
\end{figure*}

  We calculated two model grids; one with a mass-loss rate of $ 10^{-7} $ $ M_{\odot}\, \mathrm{yr}^{-1}$ and 
another with $ 10^{-8} $ $ M_{\odot}\, \mathrm{yr}^{-1}$. The mass-loss rate is scaled 
proportional to $L^{3/4}$ to preserve the emission of recombination 
lines such as H$\alpha$ and  \ion{He}{ii}\,4686. 
  In order to best fit the UV P-Cygni profiles and H$\alpha$ line, we also 
calculated models with different mass-loss rates for individual stars when 
necessary. The primary diagnostic lines used for the mass-loss rate 
determination in the UV are the resonance doublets 
\ion{N}{v\,$\lambda\lambda$1238--1242} and 
\ion{C}{iv\,$\lambda\lambda$1548--1551} (HST/ IUE range). To reproduce the \ion{N}{v\,$\lambda\lambda$1238--1242} lines, we accounted for shock generated X-ray emission in the model. For objects that were 
observed by FUSE, we made use of \ion{P}{v\,$\lambda\lambda$1118--1128}, 
\ion{C}{iv} at 1169\,\AA, and \ion{C}{iii} at 1176\,\AA. For stars with 
available UV spectra (N206-FS\,180, 187 and 214), the error in log $\dot{M}$ is 
approximately 0.1 dex. For the remaining stars, we derived the mass-loss rates 
solely on the basis of H$\alpha$ and \ion{He}{ii}\,4686. If H$\alpha$ is in 
strong absorption (no wind contamination), the mass-loss rates are based on 
nitrogen emission (\ion{N}{iii} and \ion{N}{iv}) lines with a large uncertainty. The UV P-Cygni profiles  
\ion{P}{v\,$\lambda\lambda$1118--1128}, \ion{N}{iv\,$\lambda1718$}, and 
\ion{O}{v\,$\lambda1371$} provide diagnostics for stellar wind clumping in early-type
O stars \citep{Bouret2003,Martins2011,Surlan2013}. We constrained the values of 
clumping parameters $D$ and $R_{\mathrm{D}}$ (also clumping onset) by consistently fitting these lines 
in comparison to optical lines.

  \changed{The terminal velocity $\varv_\infty$ defines the absorption trough of P-Cygni profiles in the UV}. In contrast to the optical, UV P Cygni lines are rather insensitive to temperature and abundance variations if saturated \citep{Crowther2002}. 
For stars with UV spectra, we inferred the terminal velocities from these 
P-Cygni profiles and recalculated the models accordingly.  The main diagnostic 
lines used are P-Cygni \ion{N}{v\,$\lambda\lambda$1238--1242} and 
\ion{C}{iv\,$\lambda\lambda$1548--1551}, \ion{P}{v\,$\lambda\lambda$1118--1128}, 
and \ion{S}{v\,$\lambda\lambda$1122--1134} profiles. The terminal velocity has 
been measured from the blue edge of the absorption component. The typical 
uncertainty for $\varv_\infty$ is $\pm$\,100 km\,s$^{-1}$. We calculated the 
terminal velocities theoretically from the escape velocity 
$\varv_{\mathrm{esc}}$ for those stars for which we only have optical spectra. For 
Galactic stars, the ratio of the terminal and escape velocity has been obtained 
from both theory and observations. For stars with $T_\ast\geq$\,27\,kK, the ratio 
is $\varv_\infty$/$\varv_{\mathrm{esc}}\simeq$\,2.6, and for stars with $T_\ast 
<$\,27\,kK the ratio is $\approx$1.3 \citep{{Lamers1995},{Vink2001}}. The terminal 
velocity also depends on metallicity, $\varv_\infty \, \propto 
(Z/Z_{\odot})^{q}$, where $q=0.13$ \citep{Leitherer1992}. We used this scaling 
to account for the LMC metallicity.

 We calculated our models with typical LMC abundances \citep{Trundle2007}. The mass fractions of C, 
N, and O are varied for individual objects when necessary to best fit their 
observed spectra. The abundances of these elements were determined from the 
overall strengths of their lines with uncertainties of $ \sim $20--50\,\%. The 
carbon and oxygen abundance were primarily based on \ion{C}{iii} and \ion{O}{ii} 
absorption lines. For the determination of nitrogen abundance, we particularly 
used the \ion{N}{iii} absorption lines at $\lambda$4510--4525 and \ion{N}{v} 
absorption lines.

\begin{table*}
\caption{Main diagnostics used in our spectral fitting process.} 
\label{table:lines}      
\centering

\begin{tabular}{ccc}
\hline
\hline
\noalign{\vspace{1mm}}
Parameter  & UV lines  & Optical lines  \\
\noalign{\vspace{1mm}}
\hline 
\noalign{\vspace{1mm}}
$T  _\ast$     &  &\ion{He}{i}/\ion{He}{ii}, \ion{N}{iii}/ \ion{N}{iv}/ 
\ion{N}{v} line ratios.\\
\noalign{\vspace{1mm}}
\hline 
\noalign{\vspace{1mm}}
log\,$g_\ast$       &  &H$\gamma$ and H$\delta$ line wings \\
\noalign{\vspace{1mm}}
\hline 
\noalign{\vspace{1mm}}
$\dot{M}$   &  \ion{N}{v\,1238--1242}, \ion{C}{iv\,1548--1551}  & H$\alpha$, 
\ion{He}{ii}\,4686 emission\\
 & \ion{N}{iv}\,1718, \ion{O}{iv\,1338--1343}, \ion{He}{ii}\,1640\\
 & \ion{P}{v\,1118--1128}, \ion{C}{iv} 1169, \ion{C}{iii} 1176 &\\
 \noalign{\vspace{1mm}}
\hline 
\noalign{\vspace{1mm}}
$\varv_\infty$ &   \ion{N}{v\,1238--1242}, \ion{C}{iv\,1548--1551}& H$\alpha$ line
width\\
 &  \ion{Si}{iv\,1393--1403}, \ion{N}{iv}\,1718 \\
 & \ion{P}{v\,1118--1128}, \ion{C}{iv} 1169, \ion{C}{iii} 1176  blue edge&\\
 \noalign{\vspace{1mm}}
\hline 
 \noalign{\vspace{1mm}}
$D$ &\ion{P}{v\,1118--1128}, \ion{N}{iv\,1718}, \ion{O}{v\,1371}& H$\alpha$\\
\noalign{\vspace{1mm}}
\hline 
\noalign{\vspace{1mm}}
$\varv\,\sin i$&  & \ion{N}{iii\,4640--4642}, \ion{N}{iv\,4058}\\
&  & \ion{Si}{iv\,4089--4116}, \ion{He}{i}~4388\\
\noalign{\vspace{1mm}}
\hline 
\noalign{\vspace{1mm}}
Abundances     &  &\ion{C}{iii\,4647--4650--4651}, \ion{Si}{iv\,4089--4116}\\
   &  &\ion{O}{ii\,4649}, \ion{O}{ii\,4070}, \ion{N}{iii\,4510--4525}, \ion{N}{v}\,4620\\
\noalign{\vspace{1mm}}
\hline 
\noalign{\vspace{1mm}}
\end{tabular}

\end{table*}

Finally, the projected rotation velocity $\varv\,\sin i$ is constrained from the 
line profile shapes.  We used the \texttt{iacob-broad} tool implemented in IDL, 
which was developed by \citet{Simon-diaz2014}. This tool provides $\varv$\,sin\,$i$ values 
based on a combined Fourier transform (FT) and goodness-of-fit (GOF) analysis. 
We used two of the methods described in \citet{Simon-diaz2014}.
 In the first method, the $\varv\,\sin i$ comes from the first zero of the 
FT method. In this case the assumption is that 
macroturbulence is negligible, i.e., the additional broadening is by rotation.

 The second fitting method is a combination of FT and GOF analysis, where the 
line profile is additionally convolved with a macroturbulent profile. The 
macroturbulent velocity ($\varv_{\mathrm{mac}}$) is calculated from the GOF when 
 $\varv\,\sin i$ is fixed to the value corresponding to the first zero of the 
FT.  
 
 We selected \ion{He}{i}, \ion{Si}{iv} absorption lines and \ion{N}{iii} and 
\ion{N}{iv} emission lines for determining $\varv\,\sin i$ using these methods. In some cases, absorption lines which are not pressure broadened (e.g., He I and metal lines) are very weak or completely absent in the spectra. In this case, we used the \texttt{iacob-broad} tool on emission lines that form close to the photosphere, such as \ion{N}{iv\,4058} and \ion{N}{iii\,4510--4525}. To ensure that the inferred values of $\varv\,\sin i$ from these emission lines are reliable, we calculated the synthetic spectrum via a 3D integration algorithm that accounts for rotation and that is appropriate for nonphotospheric lines \citep{Shenar2014}. These calculations resulted in spectra which are similar to the convolved ones. Hence, the \texttt{iacob-broad} tool is valid for these emission lines. For binaries, we selected those lines which have only contribution from either a primary or secondary. The fitted line profiles for individual stars are given in Fig\,\ref{fig:rotaionprof}. The formal uncertainty in $\varv\,\sin i$ is $ \sim$\,3--8\,km\,s$ ^{-1} $. Corresponding to these velocities, the model spectra are convolved with rotational and macroturbulent profiles. This yields consistent 
fits with the observations.

The main diagnostic lines for stellar and wind parameters used in our spectral 
fitting process are summarized in Table\,\ref{table:lines}. 

The luminosity and the color excess $E _{B-V} $ of individual objects were 
determined by fitting the model SED to the photometry. The model flux is scaled 
with the LMC distance modulus of 18.5\,mag, which corresponds to a distance of 
50 kpc \citep{Pietrzynski2013}. The uncertainty in the luminosity determination 
is an error propagation from color excess (relatively small for LMC stars), 
temperature, and observed photometry. All these uncertainties give a final 
accuracy of about 0.2 in log $L/L_{\odot}$. For stars with available 
flux-calibrated UV spectra (HST, IUE, or FUSE), the model SED is iteratively 
fitted to this calibrated spectra and normalized consistently by dividing the reddened model 
continuum. The uncertainty in luminosity is smaller in these cases ($\approx$0.1 dex).

 Subsequently, individual models with refined stellar parameters and abundances 
are calculated for each Of-type star in our sample. All these fitting processes were performed iteratively until no further improvement of the fit was possible. 
 
 \subsubsection{Single star model}
 \label{subsubsec:single}

\begin{figure*}
\centering
\includegraphics[scale=0.8]{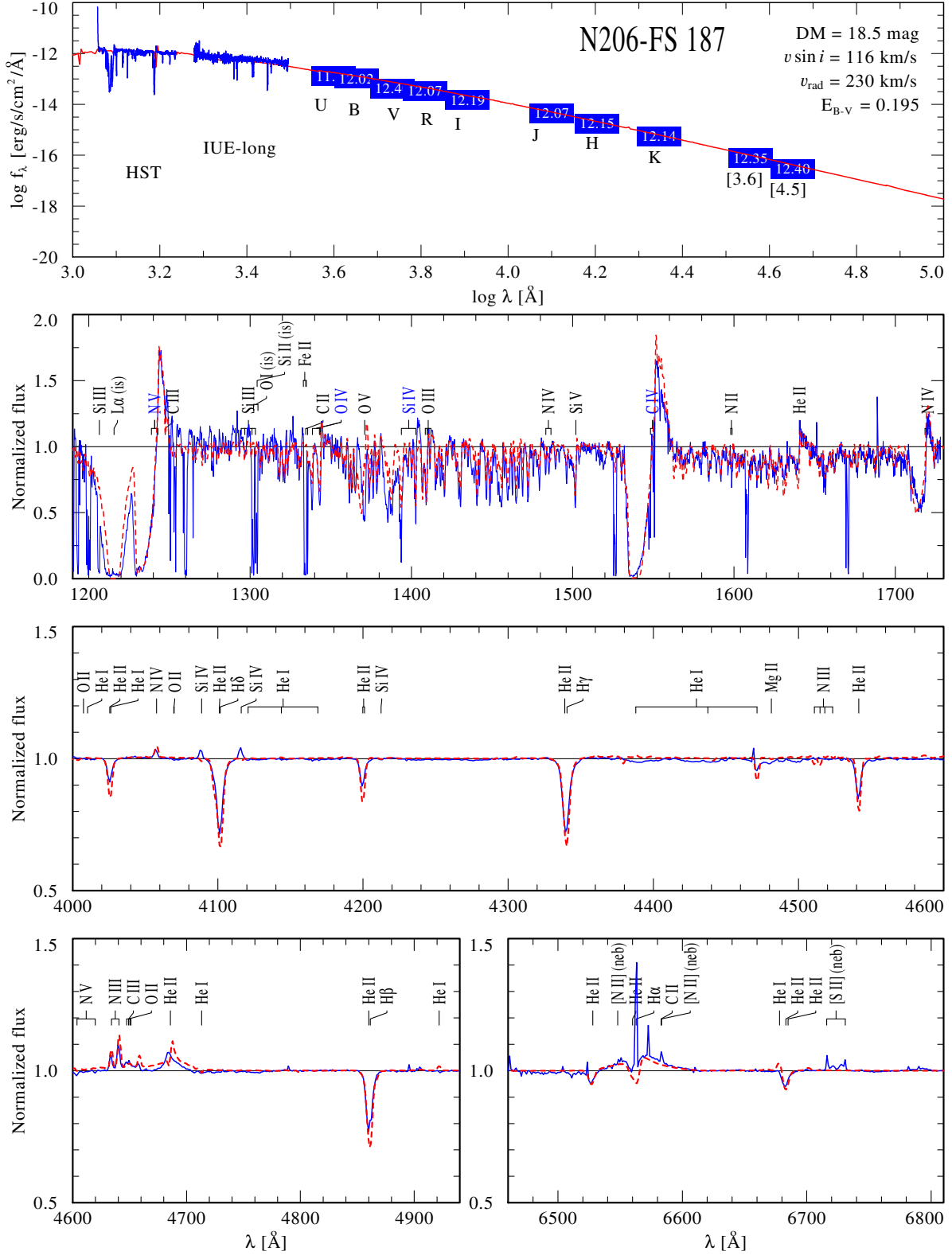}
\caption{Spectral fit for N206-FS\,187. The upper panel shows the model SED 
(red) fitted to the available photometry from optical ($UBV$ and $I$) and infrared 
($JHK_{s}$ and $IRAC$ 3.6 \& 4.5 $\rm \mu m$) bands (blue boxes) as well as the calibrated 
UV spectra from HST and IUE. The lower panels show the normalized HST and 
VLT-FLAMES spectra (blue solid line), overplotted with the PoWR model (red 
dashed line). The parameters of this best-fit model are given in 
Table\,\ref{table:stellarparameters}. The observed spectrum also contains 
nebular emission lines (neb) and interstellar absorption lines (is).}
\label{fig:RHH187}
\end{figure*}

As an example, the fit for N206-FS\,187 is given in Fig.\,\ref{fig:RHH187}. The 
upper panel of the figure shows the theoretical SED fitted to multiband photometry and calibrated UV spectra. We appropriately varied the reddening and luminosity  for fitting the  observed data with the 
model SED. Reddening includes the contribution from the Galactic foreground 
($E_{B-V} $ = 0.04\,mag) adopting the reddening law from \citet{Seaton1979}, and 
from the LMC using the reddening law described in \citet{Howarth1983} with 
$R_{V}=3.2$. The total $E_{B-V} $ is a fitting parameter. Since flux calibrated 
HST/STIS and IUE spectra are available for this star, reddening and luminosity are 
well constrained.

The second panel shows the normalized high resolution HST/STIS UV spectrum fitted to 
the model. This spectrum was consistently normalized  with the reddened model 
continuum. The last three panels show the normalized VLT-FLAMES spectrum 
(normalized by eye) in comparison to the PoWR spectrum.

 This star is best fitted with a model of $T  _\ast$ = 38 kK, based on the 
\ion{N}{iii}\,/ \ion{N}{iv} line ratio. The Balmer absorption lines in the 
observation are weaker compared to the model, because they are partially filled with nebular emission. The broad emission in H$\alpha$ and 
\ion{He}{ii}\,4686 is formed in the stellar wind, which provides a constraint 
to the mass-loss rate. 
  
 \begin{figure}
\includegraphics[scale=0.52,trim={3cm 16.8cm 1cm 5cm}]{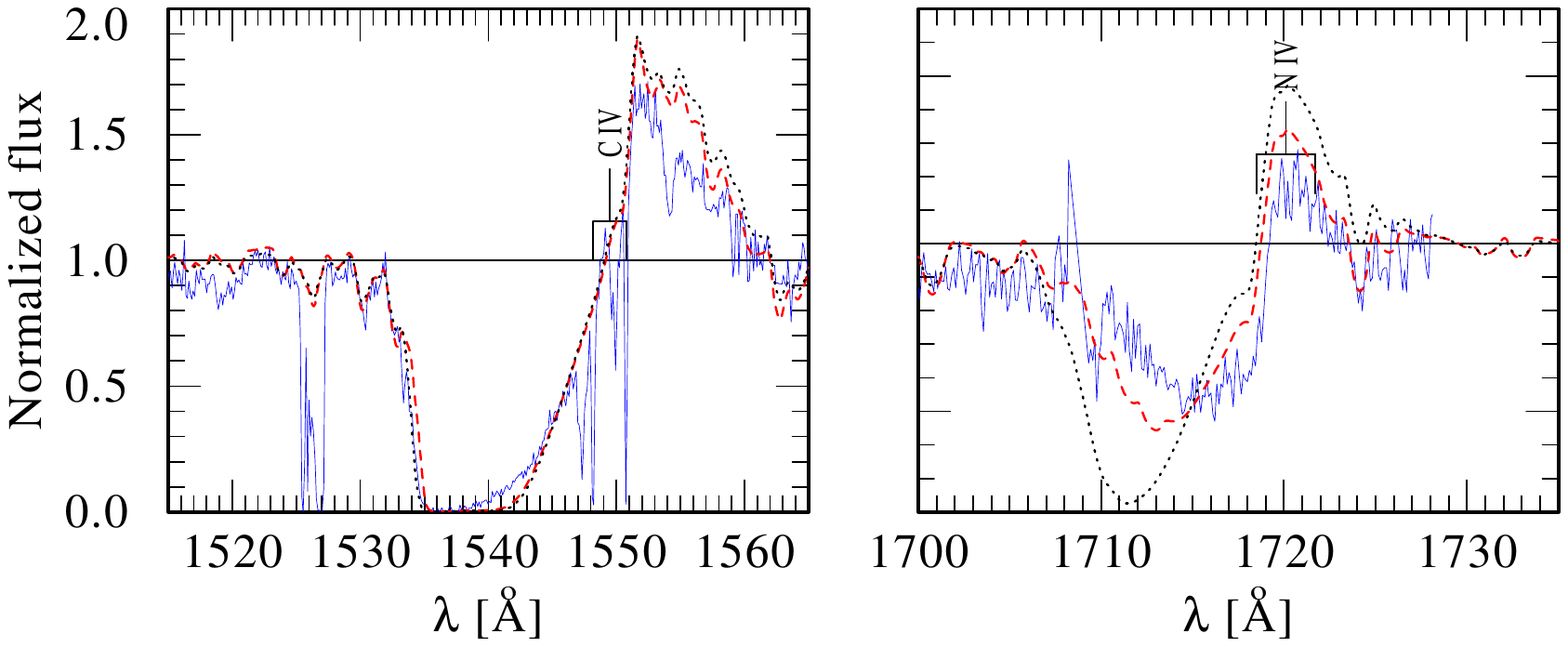}
\caption{Models with different clumping parameters are compared to the 
P-Cygni profiles \ion{C}{iv\,$\lambda\lambda$1548--1551} and 
\ion{N}{iv\,$\lambda1718$} of N206-FS\,187 (blue solid line). Dashed lines (red) 
are for the model with $\dot{M} =2.6  \times 
10^{-6}\,M_{\odot}\,\mathrm{yr^{-1}}$, $D=20$, and $R_{\mathrm{D}} = 
0.05\,R_\ast$. The dotted lines (black) show the model with $\dot{M} =1  \times 
10^{-5}\,M_{\odot}\,\mathrm{yr^{-1}}$, $D=10$, and $R_{\mathrm{D}} = 
10\,R_\ast$.}
\label{fig:CN187}
\end{figure}
 
 The wind parameters of this star are better estimated by fitting the strong 
P-Cygni \ion{N}{v} and \ion{C}{iv} profiles. \changed{The \ion{N} {v\,$\lambda\lambda$1238--1242} line is better reproduced by incorporating the X-ray field in the model, with X-ray luminosity $L_{\rm X} =4.7 \times 10^{33}$\,erg\,s$^{-1}$.} The X-ray field influences the ionization structure and especially strengthen this \ion{N}{v} line \citep{Cassinelli1979,Baum1992}. 
Moreover, this star is detected as a X-ray point source in \emph{XMM-Newton} 
observation of \citet{Kavanagh2012}. Its strong X-ray luminosity might be due to binarity (see Sect.\,\ref{sect:xray}), but we do not see any sign of secondary component in the spectra. Therefore, the star is here fitted as a single star.

The left panel of Fig.\,\ref{fig:CN187} 
shows the \ion{C}{iv\,$\lambda\lambda$1548--1551} resonance doublet. We measured 
$\varv_\infty$ from its blue edge as $2300 \pm 50$ km\,s$^{-1}$. Since this line 
is saturated, it is not sensitive to clumping. But the \ion{N}{iv\,$\lambda1718$} line shown in the right panel is unsaturated, and found to be weaker in the observed spectrum than in the model. This could be an indication of strongly clumped wind. Figure\,\ref{fig:CN187} compares two models with different clumping parameters as described in Sect.\,\ref{subsect:models}. In 
order to decrease the strength of the \ion{N}{iv\,$\lambda1718$} P-Cygni profile 
in the model, clumping should start before the sonic point and reach its maximum 
value D already at a very low radius. The model with a density contrast $D=20$ and the clumping starts at a radius of 0.025$\,R_\ast$ and reaches the maximum value of $D$ at $R_{\mathrm{D}} = 0.05\,R_\ast$, best reproduces the observation (see right panel of Fig.\,\ref{fig:CN187} ). Since clumping enhances the emission of H$\alpha$, 
we decreased the mass-loss rate by a factor of 4 compared to the model with the
default clumping. The \ion{O}{v\,$\lambda1371$} line is similarly affected by 
wind clumping. However, the strong X-ray luminosity and possible binarity (see Sect.\,\ref{sect:xray}) of this source might be related to the strength of these lines.

\subsubsection{Composite model}
 \label{subsubsec:composite}

The spectrum of N206-FS\,187 shown in Fig.\,\ref{fig:RHH187} is satisfactorily fitted by a single-star model. However, there are two objects in our sample (N206-FS\,180, N206-FS\,131) for which the observed spectra cannot be reproduced by a single synthetic spectrum.

The spectrum of N206-FS\,180 is shown in Fig.\,\ref{fig:RHH180}.
The weakness of the \ion{N}{iii} lines, \ion{N}{iv} lines in emission, and the 
presence of strong \ion{N}{v} absorption lines  indicate a very high stellar 
temperature. We selected the temperature of the model so that the synthetic 
spectrum reproduces the observed \ion{N}{iv}/\ion{N}{v} and 
\ion{N}{iii}/\ion{N}{iv} line ratios. The best fit is achieved for $T_\ast$ = 
50\,kK and log\,$g_\ast$ = 4.2. Since the temperature of this model is very 
high, it does not predict any \ion{He}{i} lines to show up. Hence, the small 
\ion{He}{i} absorption lines in the observed spectrum can be attributed to a 
companion, which must be an O star of late subtype (later than O6 to 
produce \ion{He}{i} absorption lines, but earlier than B0 since Si, C, O, and Mg 
absorption lines are absent).

The observed spectrum is fully reproduced with a composite model (see 
Fig\,\ref{fig:BinRHH180}). The final model spectrum is the sum of an early-type 
Of star O2 V((f*)) and an O8-9 star. The light ratio of the components is 
constrained by the diluted strength of absorption features which are attributed 
to the secondary component \citep{Shenar2016}. These absorption features 
should be insensitive to the remaining stellar parameters. In our case of 
N206-FS\,180, the \ion{He}{i} lines at 4388\AA,  4472\AA, and 4922\AA\, are 
available for this purpose. Here our main assumption is that the primary 
component is more luminous and hotter than the secondary. We also ensured that 
the primary model perfectly reproduces the \ion{N}{v} and \ion{N}{iv} lines, 
assuming that they have no contribution from the secondary component (late O-type). The 
individual and the composite SEDs are depicted in the upper panel of 
Fig\,\ref{fig:BinRHH180}.

 The luminosity and reddening derived from fitting IUE-short spectrum and 
photometric data are different. The UV spectrum fit best with a model SED of 
high luminosity log\,$L/L_{\odot}$\,=\,6.47, while the photometry is best fitted with a SED having
log\,$L/L_{\odot}$\,=\,6.11 (primary). Both the high resolution and low 
resolution IUE spectrum (flux calibrated) show deviation from the photometry. 
Since the aperture of the IUE is large ($\approx 21\arcsec \times 9\arcsec$) and 
the star belongs to the crowded region of the cluster, the observed spectrum is 
likely to be contaminated from nearby stars. So, we have adapted the luminosity 
corresponding to photometry.

For those stars which show H$\alpha$ in absorption, the mass-loss rate and 
terminal velocity can be solely derived from UV spectra. For N206-FS\,180, the 
normalized IUE spectrum is shown in the second panel of Fig.\,\ref{fig:RHH180}. 
The P-Cygni \ion{O}{v} line at 1371\AA\, is stronger in the model than in the 
observation, which usually indicates strong wind clumping. But according to 
\citet{Massey2005} if H$\alpha$ is in absorption, the only wind contribution is 
from layers very close to the sonic point, so absorption-type profiles are 
hardly affected by clumping. Apart from this, we could not fit the observed spectrum using a model with stronger clumping prescriptions. Hence, we consider only the default clumping with 
$D=10$ in the model (see Sect.\,\ref{subsect:models}).

\begin{figure}[!ht]
\begin{center}
\includegraphics[scale=0.8,trim={0.5cm 19.7cm 10cm 1.5cm}]{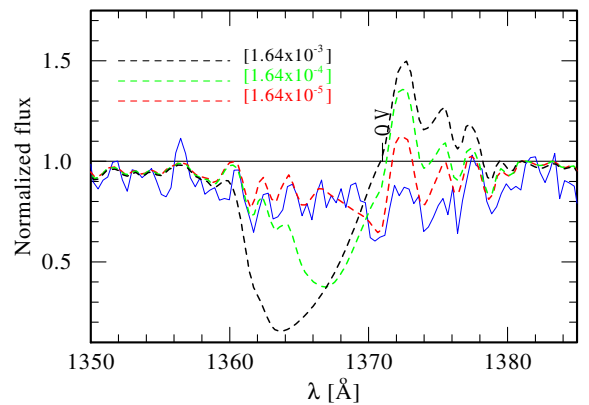}
\end{center}
\caption{The  \ion{O}{v} P-Cygni line profile of N206-FS\,180 (blue solid line). 
Three models with different oxygen mass fractions (see labels) are shown for 
comparison.}
\label{fig:CNO180}
\end{figure}

Since this star is still in the hydrogen burning phase, the increase in nitrogen abundance would result in a lower oxygen abundance (CNO balance). Models with three different oxygen mass fractions (varied by a factor of 10) are compared to the \ion{O}{v} line in 
Fig.\,\ref{fig:CNO180}. We can only reproduce this line with a model with 
100 times lower oxygen abundance compared to the typical LMC value. However, the total CNO abundance of this model is inconsistent 
with the typical LMC values.

The nitrogen lines of \ion{N}{v} and \ion{N}{iv} in the optical are also found to be affected 
by changing the mass-loss rate and terminal velocity in the model. We estimate 
$\dot{M} = 7 \times 10^{-6}\,M _{\odot}\,\mathrm{yr^{-1}} $ for the primary from 
the best fit. The terminal velocity measured from the blue edge of 
\ion{C}{iv\,$\lambda\lambda$1548--1551} is 2800\,km\,s$ ^{-1} $. In the cases 
where UV spectra are not available and H$\alpha$ is in absorption, $\dot{M}$ has a large uncertainty and $\varv_\infty$ is theoretically determined as 
described in Sect.\,\ref{subsec:specfit}.

We suspect the star N206-FS\,131 also to be a binary. The remaining Of 
stars are well fitted with single star models. Individual descriptions and 
spectral fits of all stars of our sample are given in Appendix A and Appendix B, respectively.


\section{Results and discussions}
\label{sect:results}

\subsection{Stellar parameters}
\label{sect:sparameters}


\begin{table*}
\caption{Stellar parameters of nine Of-type stars in N\,206 superbubble.}
\label{table:stellarparameters}      
\renewcommand{\tabcolsep}{3.5pt}

\begin{center}
\begin{tabular}{lcccS[table-format = 
<-1.3]ccrcS[table-format=3.1]S[table-format=3.1]ccl}
\hline 
\hline
\noalign{\vspace{1mm}}
N206-FS  &	$T  _\ast$ 	& log\,$L$ &	log\,$g_\ast$ &	
\multicolumn{1}{c}{log $\dot{M}$} &	$E_{B-V} $&	$M_{\mathrm{V}}$ 	
&	\multicolumn{1}{c}{$R _\ast$}  &	
\multicolumn{1}{c}{$\varv_\infty$} 	&\multicolumn{1}{c}{ $\varv$\,sin\,$i$}  
&	\multicolumn{1}{c}{$M  _\ast$\tablefootmark{(1)}} &	log $Q$ &	
log $L_{\mathrm{mec}}$\tablefootmark{(4)}& Spectral type\\ 

\# & [kK] & [$L _{\odot}$]&[cm s$ ^{-2} $]& [$M _{\odot}\,\mathrm{yr}^{-1} $] & 
[mag] &[mag]& [$R _{\odot}$] & 
\multicolumn{1}{c}{[km\,s$^{-1}$]}&[km\,s$^{-1}$]&[$M _{\odot} $]&[s$ ^{-1} $] 
&[$L _{\odot}$] &\\
 
\noalign{\vspace{1mm}}

\hline 
\noalign{\vspace{1mm}}
66 	                        & 35.0 	& 5.22 	& 4.0 	& -6.38 	   & 
0.17 	& -4.61 	& 11.1 		  & 2200 	& 67 	& 45 	& 48.7 	
& 2.22 	&O8 IV((f))  \\
111 	                    & 37.0 	& 5.03 	& 4.2 	& -6.85 	   & 
0.10 	& -4.20 	& 8.0 	  & 2500 	& 43 	& 37 	& 48.6 	& 1.86 	
&O7 V((f))z \\
131a\rlap{\tablefootmark{(2)}} 	& 36.0 	& 5.80 	& 3.6 	& -5.50  	   & 
0.11 	& -6.23 	& 20.5 		  & 1500 	& 96 	& 61 	& 49.4 	
& 2.77 &	O6.5 II(f)  \\
131b\rlap{\tablefootmark{(3)}} 	& 30.0 	& 5.04 	& 3.8 	& -6.57 	   & 
0.12 	& -5.00 	& 12.3 	  & 1900 	& 79 	& 35 	& 48.0 	& 1.91  
& O9.7 IV\\
162                   	    & 34.0 	& 5.60 	& 3.8 	& -6.03     & 0.20 	
& -5.67 	& 18.2 	  & 2100 	& 83 	& 77 	& 49.1 	& 2.53 	&O8 IV((f))e \\
178                   	    & 35.0 	& 5.22 	& 4.0 	& -6.38     & 0.14 	
& -4.55 	& 11.1 		  & 2200 	& 79 	& 45 	& 48.7 	& 2.22 
&O7.5 III((f))\\
180a\rlap{\tablefootmark{(2)}} 	& 50.0 	& 6.11 	& 4.2 	& -5.32 	   & 
0.16 	& -5.77 	& 15.2 		  & 2800\rlap{\tablefootmark{(7)}}	
& 128 	& 133   & 50.0 	& 3.49&O2 V((f*))\\
180b\rlap{\tablefootmark{(3)}} 	& 32.0 	& 4.91 	& 4.2 	& -6.62 	   & 
0.18 	& -4.50 	& 9.3 		  & 2900 	& 42 	& 50 	& 48.0 	
& 2.21 & O8 IV\\
187\rlap{\tablefootmark{(5)}} 	    & 38.0 	& 6.28 	& 3.6 	& -4.80 & 0.19 	
& -6.70 	& 31.9  & 2300\rlap{\tablefootmark{(7)}}	& 116 	& 148 	
& 50.0 	& 3.86  &O4 If  \\
193 	                    & 36.0 	& 5.20 	& 4.2 	& -6.36 	   & 
0.20 	& -4.44 	& 10.3 		  & 2900 	& 103 	& 61 	& 48.7 	
& 2.48 	&O7 V((f))z  \\
214\rlap{\tablefootmark{(6)}}   	& 38.0 	& 6.10 	& 3.6 	& -4.93 & 0.33 	
& -6.26 	& 26.0 	& 2300\rlap{\tablefootmark{(7)}} 	& 98 	& 98 	
& 49.8 	& 3.71 &O4 If \\

\hline 
\end{tabular}
\end{center}
\tablefoot{
\tablefoottext{1}{Spectroscopic masses}
\tablefoottext{2}{Primary component}
\tablefoottext{3}{Secondary component}
\tablefoottext{4}{The mechanical luminosity $0.5 \dot{M} \varv_\infty^{2}$}
\tablefoottext{5}{When strong clumping ($D=20$ and $R_{\mathrm{D}} = 
\,0.05\,R_\ast$) is adopted then log $\dot{M} = -5.59\,M _{\odot}\, 
\mathrm{yr}^{-1}$ and log\,$L_{\mathrm{mec}}=3.05\,L _{\odot}$. \changed{The X-ray field is included in this model, with X-ray luminosity $L_{\rm X} = 4.7 \times 10^{33}$\,erg\,s$^{-1}$.}}
\tablefoottext{6}{When strong clumping ($D=20$ and $R_{\mathrm{D}} = 
\,0.05\,R_\ast$) is adopted then log $\dot{M}=-5.72\,M _{\odot}\, 
\mathrm{yr}^{-1}$ and log\,$L_{\mathrm{mec}}=2.92\,L _{\odot}$. \changed{The X-ray field is included in this model, with X-ray luminosity $L_{\rm X} = 1.7 \times 10^{33}$\,erg\,s$^{-1}$.}} 
\tablefoottext{7}{$\varv_\infty$ is determined from UV P-Cygni profiles. Other 
values are theoretically calculated from $\varv_{\mathrm{esc}}$.}
}
\end{table*}


 The fundamental parameters for the individual stars are given in 
Table\,\ref{table:stellarparameters}. The rate of hydrogen ionizing photons (log 
Q) and the mechanical luminosity of the stellar winds 
($L_{\mathrm{mec}}\,=\,0.5\,\dot{M}\,\varv_\infty^{2}$) are also tabulated. All 
these models are calculated with default clumping parameters as described in 
Sect.\,\ref{subsect:models}. 

The gravities shown here are not corrected for rotation. The effect is insignificant 
and much less than the uncertainty values, even for the fastest rotating star 
in our sample (log\,$(g_\ast +(\varv\,\sin\,i)^{2}/R _\ast ) - $log\,$g_\ast < 
0.03$). The reddening of our sample stars is only $0.1-0.3$\,mag (see 
Table\,\ref{table:stellarparameters}). The absolute visual magnitudes 
($M_{\mathrm{V}}$) in the table are also derived from the respective models. The 
spectroscopic masses, calculated from log\,$g_\ast$ and $R  _\ast$ 
(using relation $g_\ast=G\,M_\ast\,R_\ast^{-2}$), are in the range $30-150\,M _{\odot} $.

  For the binary candidates N206-FS\,180 and N206-FS\,131, the stellar 
parameters for the individual components (`a' denotes the primary and `b' the 
secondary component) are given in the table as well. Only the primaries of these binary systems are considered in the following discussions.

\begin{figure}[!ht]
\centering
\includegraphics[scale=0.48]{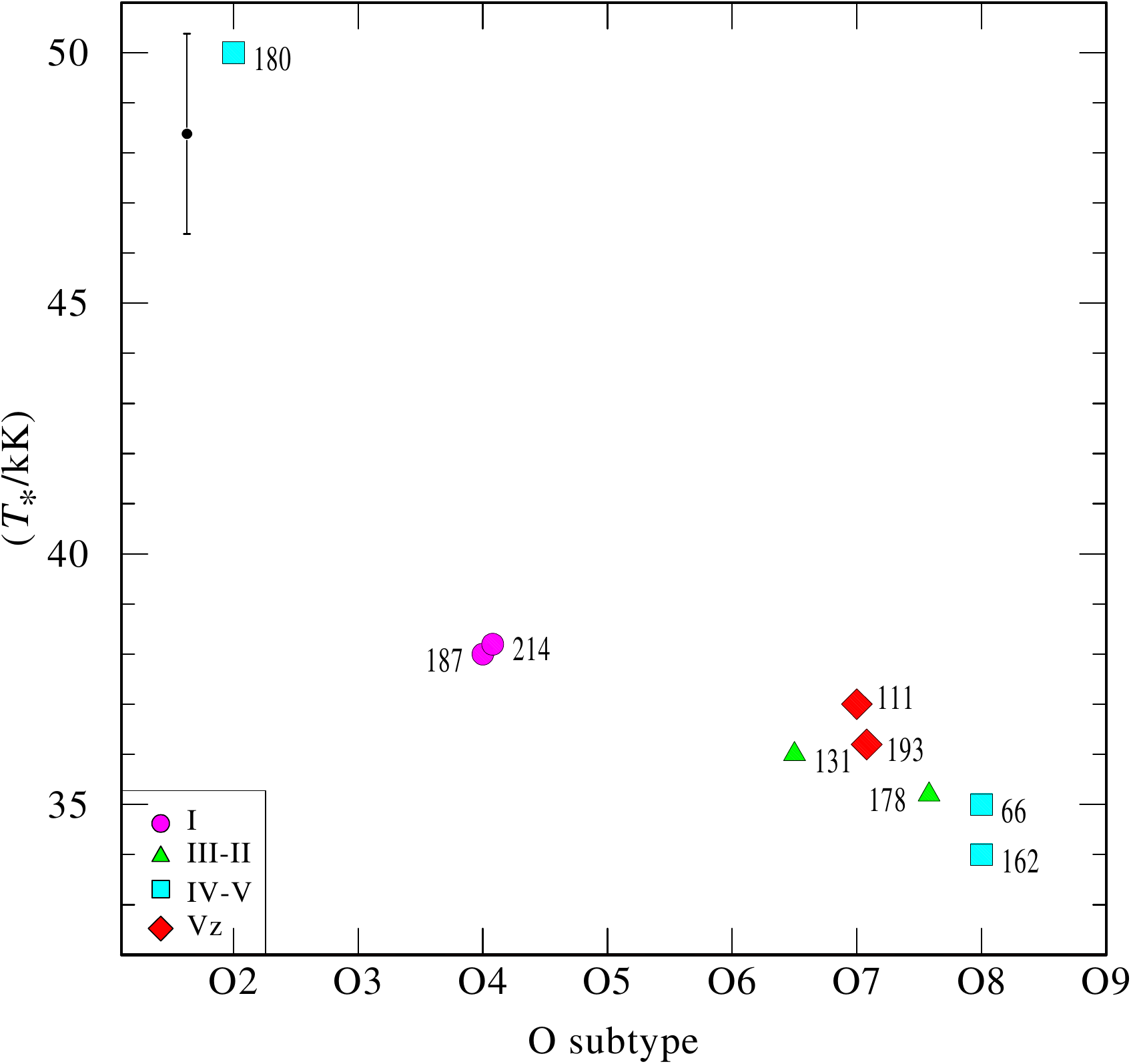}
\caption{Effective temperature as a function of spectral type. Squares, 
triangles, and circles denote the luminosity classes Vz, V-IV, III-II and I, 
respectively. Typical uncertainties are indicated by the error bar in the upper 
left corner. }
\label{fig:specT}
\end{figure}

Figure\,\ref{fig:specT} shows how the effective temperature correlates with the
spectral subtypes of our sample. Different luminosity classes are denoted using squares, 
triangles, and circles for luminosity class V-IV, III-II, and I, respectively. 
The O2 subtype shows an outstandingly high effective temperature, $T 
_\ast$\,=\,50\,kK, which is comparable to the temperatures of 
other O2 stars (e.g., \citealt{Gonzalez2012b,Gonzalez2012a,Walborn2004}). This supports the results of \citet{Mokiem2007A} and \citet{Gonzalez2012a,Gonzalez2012b}, who suggest a steeper slope in the 
temperature - spectral type relation for the earliest subtypes (O2-O3). 

%

\subsection{Wind parameters}
\label{sect:wparameters}
In addition to the stellar parameters, the spectra provide information on the 
stellar winds of our sample. The primary stellar wind parameters are wind 
terminal velocities ($\varv_\infty$) and mass-loss rates ($\dot{M}$). The 
mass-loss rate has great influence on the evolution of massive stars. The $\dot{M}$ 
scales with the metallicity of stars \citep{Leitherer1992,Vink2001}. The lower 
LMC metallicity results in less efficient wind driving in comparison to the 
Galactic environment, and consequently in lower mass-loss rates 
\citep{Vink2000}. 

For stars with available UV spectra (N206-FS\,180, N206-FS\,187, and 
N206-FS\,214), the mass-loss rate is obtained from P Cygni lines as explained in 
Sect.\,\ref{subsec:specfit}. For N206-FS\,131, H$\alpha$ is partially filled 
with wind emission, and the corresponding spectral fit yields $\dot{M}$. The rest of the sample shows 
H$\alpha$ in pure absorption, so the $\dot{M}$ values are estimated from the 
nitrogen emission lines. Since these nitrogen lines are also affected by the 
nitrogen abundance, these  mass-loss rates have an additional uncertainty. The 
derived mass-loss rates of our Of samples are in the range from 
$10^{-6.9}\,\rm{to}\, 10^{-4.8}\,$ $M _{\odot}\,\mathrm{yr}^{-1}$. 

\begin{figure}[!ht]
\centering
\includegraphics[scale=0.48]{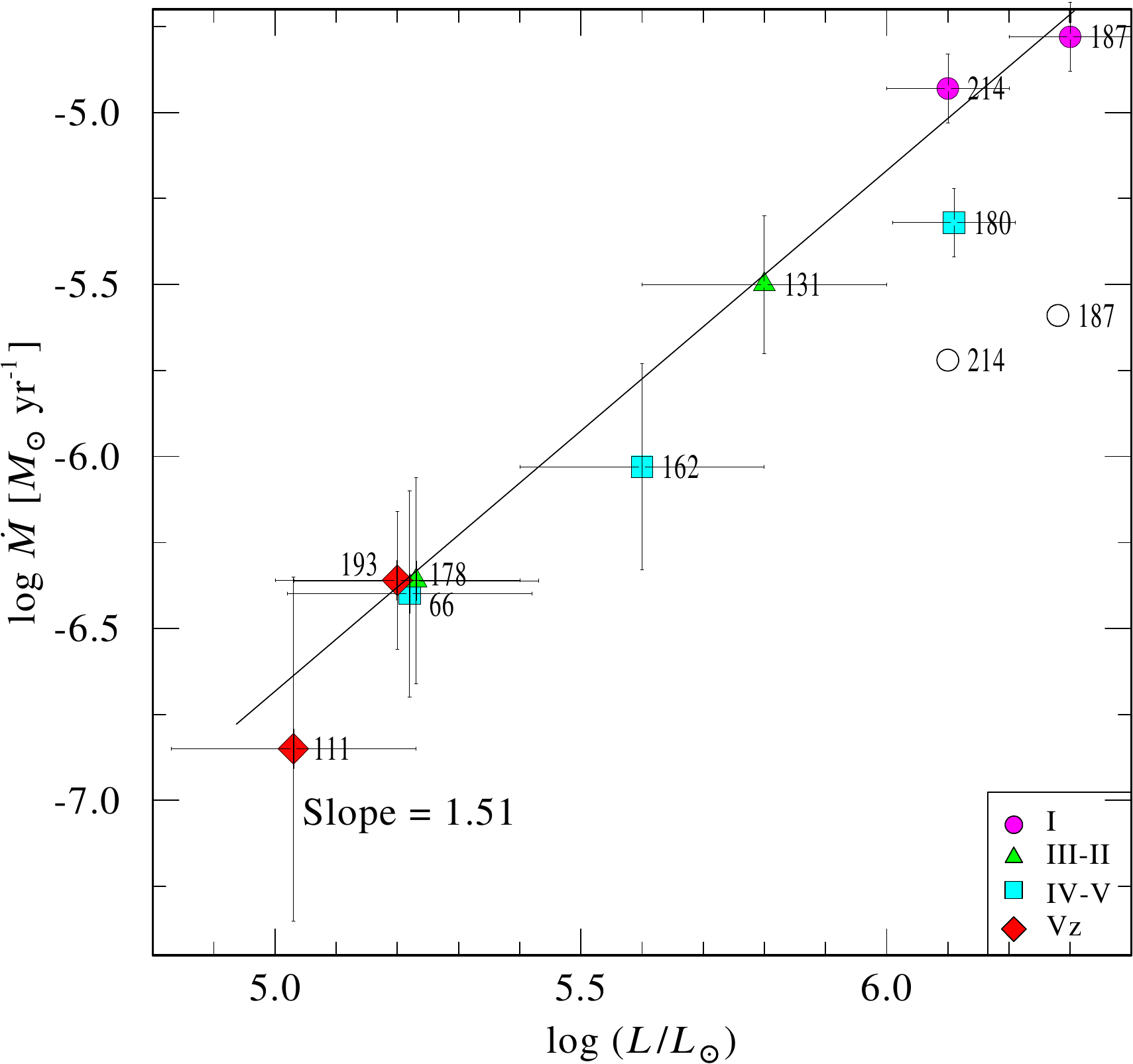}

\caption{Mass-loss rates vs. luminosity for the Of stars in the N\,206 
superbubble in the LMC. Open circles denote alternative values for the two 
supergiants obtained with stronger clumping (see Sect.\,\ref{subsubsec:single}). 
A power law is fitted to the data points. Symbols have the same meaning as in  
Fig\,\ref{fig:specT}. \changed{The error bars are shown for each source.}}
\label{fig:logL_logmdot}
\end{figure}


Mass-loss rate and luminosity exhibit a clear correlation in 
Fig.\,\ref{fig:logL_logmdot}. Here the supergiants are in the upper right corner 
with high luminosity and $\dot{M}$. \changed{This relation can be fitted as a power-law 
with an exponent $\approx 1.5$ (uncertainties are considered in the fit)}. The models of two supergiants with alternative 
clumping prescriptions (see Sect.\ref{subsubsec:single}) would yield a lower 
mass-loss rate by a factor 5 (open circles). The supergiant N206-FS\,187 has a 
very high luminosity and  mass-loss rate, even compared to the Galactic stars 
with the same spectral type. A star (VFTS\,1021) with similar high values of $L$ and $ \dot{M}$ has been found by \citet{Bestenlehner2014}. 

\subsubsection*{Wind-momentum luminosity relationship}
\label{sect:windML}

\begin{figure}[!ht]
\centering
\includegraphics[scale=0.48]{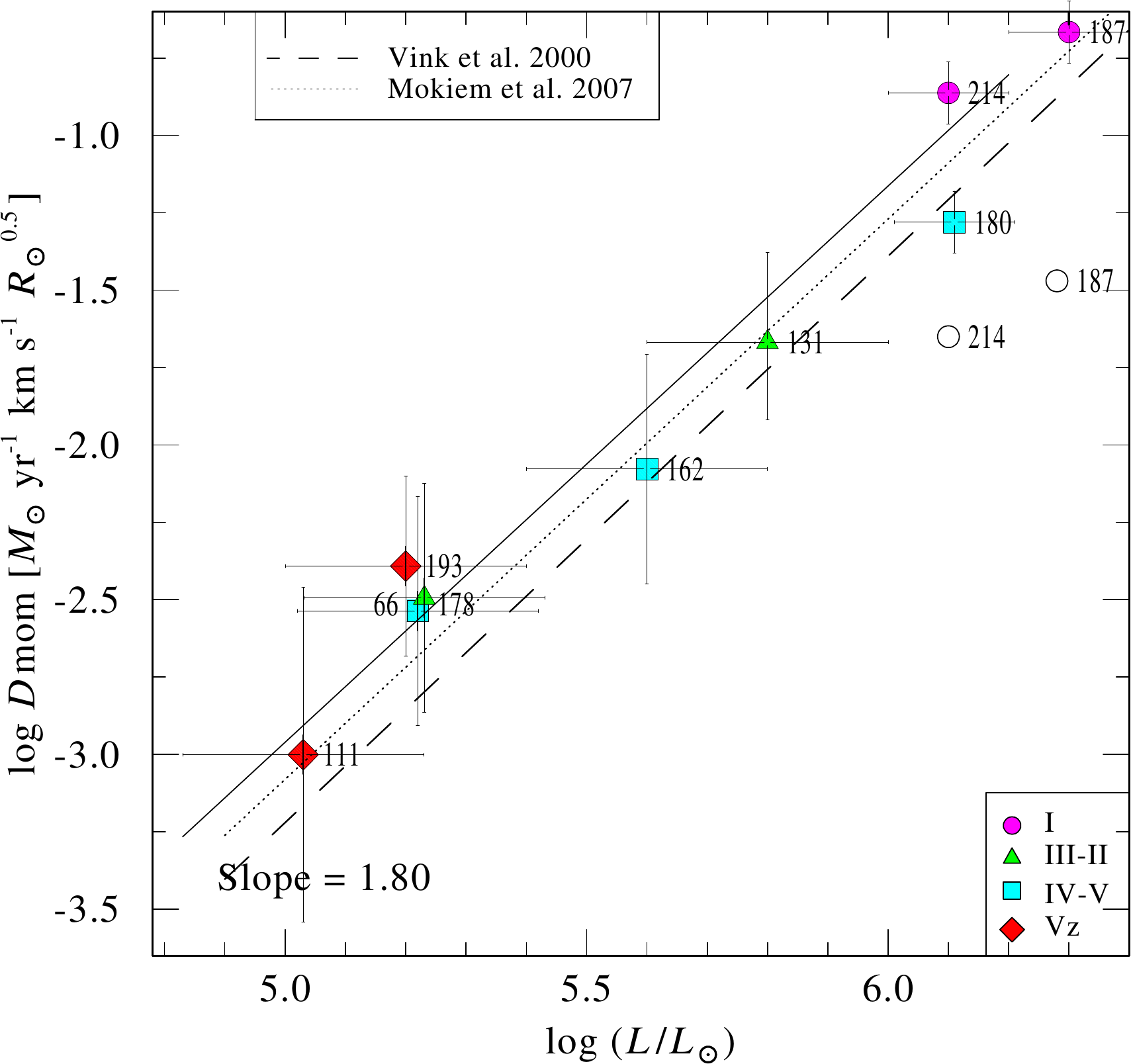}
\caption{Modified wind momentum ($D_{\mathrm{mom}}$) in units of $M 
_{\odot}\,\mathrm{yr}^{-1}$\,km\,s$^{-1}\, R_{\odot}^{0.5} $ as a function of 
the stellar luminosity for the analyzed LMC Of stars. 
The alternative values for the two supergiants adapting strong clumping (see 
Sect.\,\ref{subsubsec:single}) are indicated in open circles. The data are 
fitted by a power law (solid line). The theoretical WLR from \citet{Vink2000} 
(dashed line) and empirical WLR from \citet{Mokiem2007A} for LMC stars are also 
plotted. Luminosity classes are distinguished by different symbols. Individual error bars are given in the plot.}
\label{fig:logL_logDmom}
\end{figure}


To investigate the winds of Of stars in the LMC quantitatively, we plotted the  modified wind momentum - luminosity relation (WLR) in  
Fig.\,\ref{fig:logL_logDmom}. This diagram depicts the modified stellar wind 
momentum $ D_{\mathrm{mom}} \equiv \dot{M} \varv_\infty R_\ast^{0.5} $ \citep{Kudritzki2000}, as a function of the stellar luminosity. 

For the WLR, a relation in the form
\begin{equation}
\mathrm{log} D_{\mathrm{mom}}\, =\, x \,\mathrm{log} (L_\ast/L_{\odot})\,+\, 
\mathrm{log} D_{0}
\end{equation}
is expected, where $x$ corresponds to the inverse of the slope of the line-strength distribution function, and $D _{0} $ is related to the effective number of lines contributing to stellar wind acceleration \citep{Puls2000}. The distribution of line strengths is important to compare the observed wind strengths to the predictions of line driven wind theory.
A linear regression to the logarithmic values of the modified wind momenta obtained 
in this work (see Fig.\,\ref{fig:logL_logDmom}) yields the relation
\begin{equation}
\mathrm{log} D_{\mathrm{mom}}\, =\, (1.8 \pm 0.12)\, \mathrm{log} 
(L_\ast/L_{\odot})\,+\, (-11.94 \pm 0.71 )
\end{equation}

Figure\,\ref{fig:logL_logDmom} also shows the relation for the LMC stars ($x = 1.83$) as predicted 
by \citet{Vink2000}. The dotted lines represent the empirical WLR for LMC OB 
stars ($x=1.81$) as determined by \citet{Mokiem2007A}. \citet{Bestenlehner2014} 
also found an empirical WLR for O stars in the 30 Doradus region of the LMC  
with $x = 1.45$. \changed{The Of stars in the superbubble N\,206 shows a good correlation with both empirical and theoretical WLR. It should be noted that for the wind momentum calculation of some of the sources, we adopted  theoretical $\varv_\infty$ values in the absence of available the UV spectrum. The uncertainties of parameters are also estimated and included in the linear regression.}

The wind momenta of the supergiants would be much lower if strong clumping 
is adopted (open circles). Figure\,\ref{fig:logL_logDmom} shows the WLR with 
default clumping as described in Sect.\,\ref{subsubsec:single} for all the 
stars. \changed{This is close to the theoretical relation ($x \sim $1.8). The alternative clumping prescriptions for two supergiants (strong clumping) would result a less steeper WLR relation. \citet{Mokiem2007A} also tested clumping corrections for one supergiant and obtained a less steep relation ($x \sim $1.43). Note that the theoretical WLR is based on unclumped winds. In our analysis, a depth dependent clumping was accounted for all stars. }

\subsection{Outstanding X-ray luminosity of N206-FS\,187 points to its binarity}
\label{sect:xray}

The supergiant N206-FS\,187 was detected in X-rays with the {\em XMM-Newton} 
telescope \citep{Kavanagh2012}. Since no X-ray spectral information was presented by 
\citet{Kavanagh2012}, we adopted the X-ray flux from  ``The third 
XMM-Newton serendipitous source catalog'' \citep{Rosen2016}. Correcting for 
interstellar absorption, 
$N_{\rm H}=9\times 10^{21}$\,cm$^{-2}$  \citep{Kavanagh2012}, the 
estimated  X-ray luminosity of N206-FS\,187 is
$L_{\rm X}\approx 4\times 10^{34}$\,erg\,s$^{-1}$ in the 0.2-12.0\,keV band. 
Hence, the ratio between X-ray and bolometric luminosity is 
$\log{L_{\rm X}/L_{\rm bol}}\approx -5$.  

This is an outstandingly high X-ray luminosity for an O-type star. 
Single O-type stars in the Galaxy have significantly lower X-ray 
luminosities \citep[e.g.,][]{Oskinova2005}. The high X-ray luminosity of 
N206-FS\,187 might indicate that this object is a binary. In a massive binary, 
the bulk of X-ray emission might be produced either by collision of 
stellar winds from binary components, or by accretion of stellar wind 
if the companion is a compact object. 

Some colliding wind binaries consisting of a WR and an O-type component 
have comparably high X-ray luminosities \citep{PZ2002, Guerrero2008}.  
However our spectral analysis of N206-FS\,187 rules out a WR component. 
Most likely, the binary component in N206-FS\,187 is an O-type star of 
similar spectral type. Even in this case, it is  outstandingly X-ray luminous.  
For example, the X-ray luminosity of the O3.5If* + O3.5If* binary LS\,III+46\,11 is an 
order of magnitude lower than that of N206-FS\,187 
\citep[][and references therein]{MA2015}.\changed{ Some wind-wind colliding (WWC) feature may be present in the lines such as H$\alpha$ \citep{Moffat1989}. However, as the WWC effects are usually of the order of a few percentage \citep{Hill2000}, that are not expected to highly influence the spectral features.}

On the other hand, the X-ray luminosity of N206-FS\,187 is not in the correct range
to suspect that it harbors a compact companion. The latter systems are either
usually significantly more X-ray luminous (in case of persistent sources), 
or significantly less X-ray luminous in quiescence (in case of transients)
\citep{MN2017}. 

Therefore, on the basis of the high X-ray luminosity of N206-FS\,187 and taking into 
account the uncertainties on its X-ray spectral properties, we believe
that N206-FS\,187 is one of the X-ray brightest colliding wind binaries 
consisting of stars with similar spectral types. Interestingly the other Of supergiant N206-FS\,214 does not show any X-ray emission, even though it is an eclipsing binary of similar spectral types (see Sect.\,\ref{sect:appendixa} for more details).

\changed{Note that the X-ray luminosity derived from our model using UV spectral line fit ($L_{\rm X}\approx 4.7\times 10^{33}$\,erg\,s$^{-1}$) is an order of magnitude less than this observed X-ray luminosity (see Sec.\,\ref{subsubsec:single}). One reason could be the variability of X-ray emission if it is a WWC binary. Another possibility is that the observed X-ray luminosity might be overestimated from the interstellar absorption values. }

\subsection{The Hertzsprung-Russell diagram}
\label{sect:hrd}
%
%

\begin{figure}[b]
\centering
\includegraphics[scale=0.48]{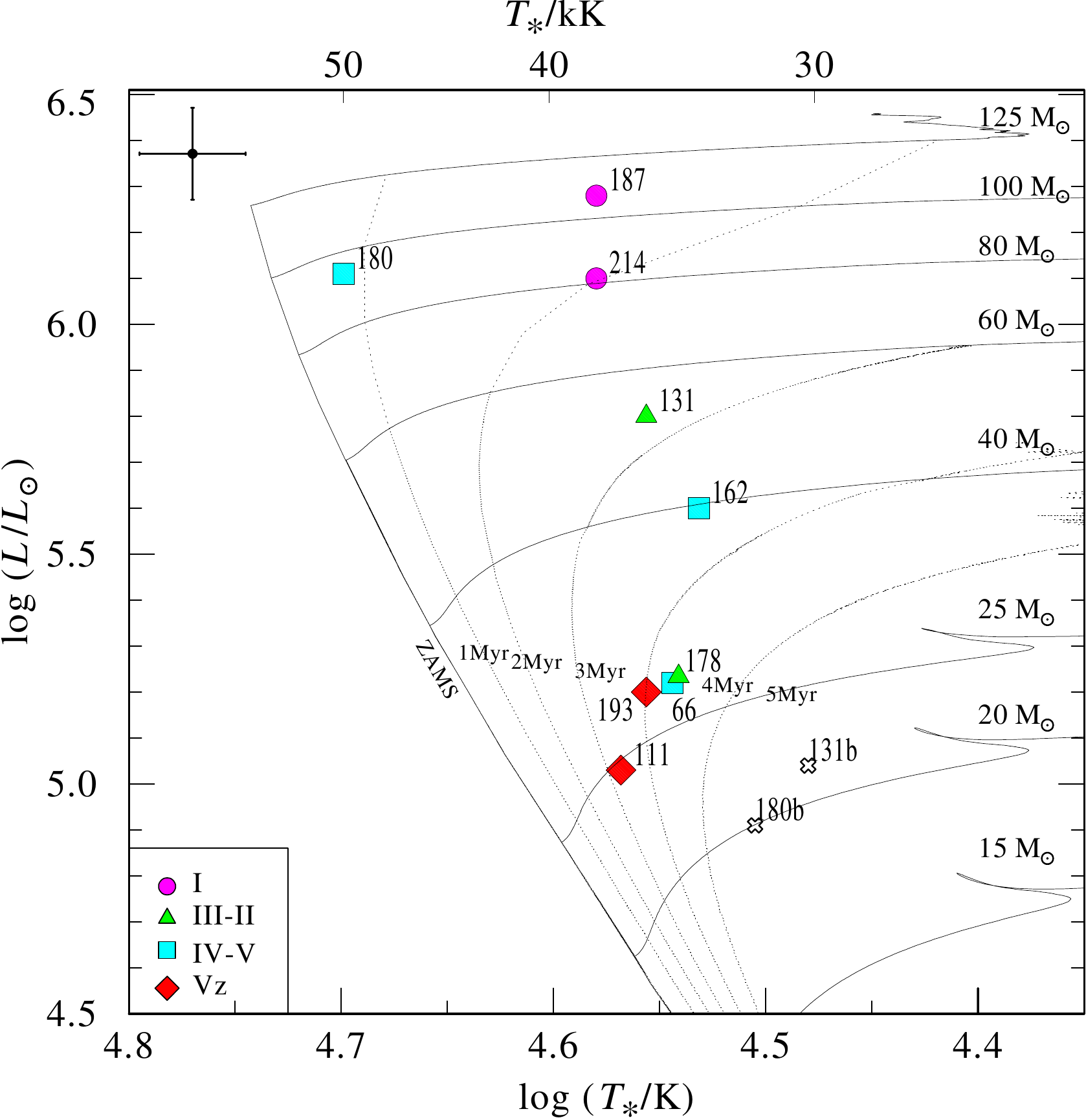}
\caption{Hertzsprung-Russell diagram for the nine Of stars in the N\,206 
superbubble in the LMC. The evolutionary tracks and isochrones are based on 
rotating ($V_{\mathrm{rot, init}} \,\sim$  100 km\,s$^{-1} $)  evolutionary 
models presented in \citet{Brott2011} and \citet{Kohler2015}. Different 
luminosity classes are denoted using rhombus, squares, triangles, and circles 
for luminosity class Vz, V-IV, III-II, and I, respectively. The position of the 
secondary components of N206-FS\,131 and N206-FS\,180 are also marked in the 
diagram with a cross symbol.}
\label{fig:hrd_of}
\end{figure}


The Hertzsprung-Russell diagram (HRD) for the Of stars in the N\,206 superbubble, 
constructed from the temperature and luminosity estimates (see 
Table\,\ref{table:stellarparameters}), is given in Fig.\,\ref{fig:hrd_of}.
The evolutionary tracks and isochrones are adapted from \citet{Brott2011} and 
\citet{Kohler2015}, accounting for an initial rotational velocity of $\sim$\,100 
km\,s$^{-1} $. The evolutionary tracks are shown for stars with initial masses 
of 15\,$ - $\,125 $M _{\odot} $, while the isochrones span from the zero age main 
sequence (ZAMS) to 5\,Myr in 1\,Myr intervals. Stars are represented by 
respective luminosity class symbols as in Fig.\,\ref{fig:hrd_of}.

The isochrones suggest that the O2((f*)) star N206-FS\,180 (primary) is very young and
close to the ZAMS (assuming single-star evolution). The evolutionary 
mass predicted from the tracks is $\approx$ 93 $M_{\odot}$ for the primary 
component of this binary system. The other Of stars are younger than $\approx$\,4\,Myr and 
more massive than $\approx$ 25 $M_{\odot}$. The supergiants in our sample are 
nearly 2 Myr old with masses in the range $70$ to $ 100\, M_{\odot}$.

\begin{figure}[!ht]
\centering
\includegraphics[scale=0.48]{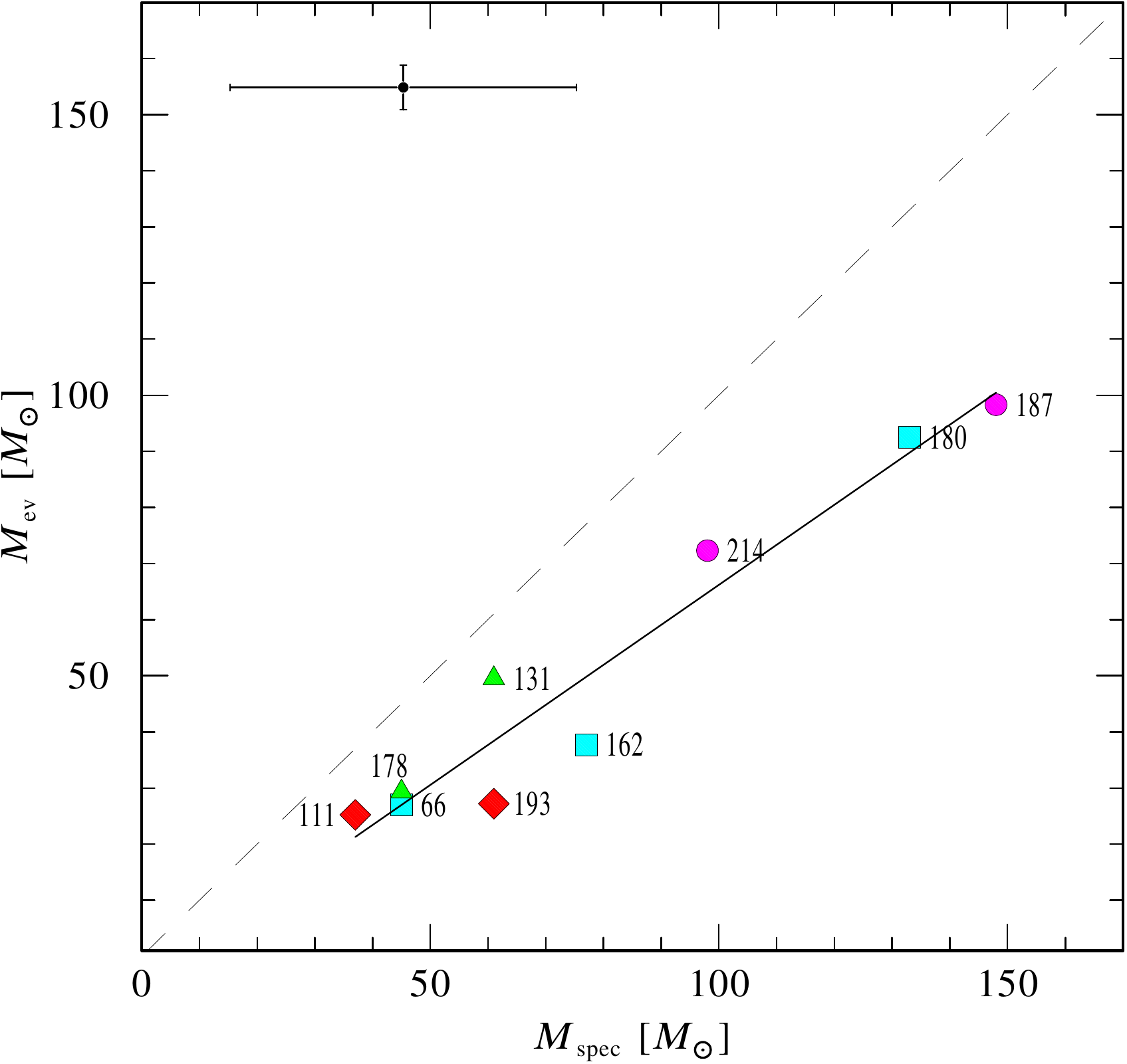}
\caption{Evolutionary masses compared to the spectroscopic masses. The 
discrepancy grows roughly linearly with the mass (see linear fit). The 
one-to-one correlation of spectroscopic and evolutionary masses is indicated by 
the dashed line.}
\label{fig:Msec_ev}
\end{figure}

The evolutionary masses from the HRD are compared in Fig.\,\ref{fig:Msec_ev} to 
the spectroscopic masses derived (see Table\,\ref{table:stellarparameters}). 
Here, the evolutionary mass refers to the current stellar mass as predicted by 
the corresponding evolutionary track. It is evident from Fig.\,\ref{fig:Msec_ev} 
that the spectroscopic masses of the objects are systematically larger than their evolutionary 
masses. However we have to acknowledge that, the errors in our spectroscopic 
mass estimates are large (at least a factor 2). The wind contamination in the Balmer wings also affects the log\,g determination, and we adopt theoretical values for $\varv_\infty$ in some cases.

This mass discrepancy problem 
was extensively investigated for Galactic and extragalactic star clusters by 
many authors \citep{Herrero1992,Vacca1996,Repolust2004}. This mass difference is 
strongly affected by the mass-loss recipe used in the evolutionary calculations 
\citep{McEvoy2015}. According to \citet{Hunter2008}, the observed 
mass discrepancy of main-sequence stars could be due to binarity, errors in the 
distance estimate, bolometric corrections, reddening, or the derived surface 
gravities. However, the systematic trend in the discrepancy may not be due to the 
uncertainties in our parameter measurements, but could originate from the evolutionary 
masses. 


\subsection{Chemical abundance}

\begin{center}
\begin{table}
\caption{Chemical abundance for the nine Of-type stars in the N\,206 superbubble 
derived from spectral fitting. The abundances are given in mass fractions.} 
\label{table:chemical}      
\centering

\begin{tabular}{cccccccccccccc}
\hline
\hline
\noalign{\vspace{1mm}}

N206-FS & X$ _{\rm N} $  	&	X$ _{\rm C} $  	&	X$ _{\rm O} $   
\\
\#& $ [10^{-3}] $ & $ [10^{-3}] $ & $ [10^{-3}] $  \\
\noalign{\vspace{1mm}}
\hline 
\noalign{\vspace{1mm}}
66& 0.28	&	0.57	&	1.64	\\
111& $ <$0.08	&	0.47&	2.64    \\
131& 0.72	&	0.23	&	1.04	 \\
162& 0.28	&	0.38	&	1.64	 \\
178& 0.7	&	0.18	&	1.64	 \\
180& 2.2	&	0.18	&	0.02	  \\
187& 0.72	&	0.23	&	3.04	 \\
193& 0.48	&	0.28	&	2.64	 \\
214& 0.72	&	0.23	&	3.04	 \\
\noalign{\vspace{1mm}}
\hline 
\noalign{\vspace{1mm}}
Typical LMC\tablefootmark{(1)} & 0.08	&	0.47	&	2.64 \\
values\\

\hline 
\end{tabular}
\tablefoot{
\tablefoottext{1}{taken from \citet{Trundle2007}}
}
\end{table}
\end{center}

The chemical abundances of our Of sample derived in terms of mass fraction are 
given in Table\,\ref{table:chemical}. These have been obtained by fitting the 
observed spectra with PoWR models calculated for different C, N, and O mass 
fractions (see Table\,\ref{table:lines} for more details). Since the nitrogen 
lines are formed by complex NLTE processes, changes in the N 
abundance can have a large impact on the strength of the lines 
\citep{Gonzalez2012a}. The P, Mg, Si, and S abundances are fixed for all the 
models. We could not reproduce the \ion{Si}{iv} emission lines in supergiants 
N206-FS\,187 and N206-FS\,214 with our models.

\begin{figure}[b]
\centering
\includegraphics[scale=0.48]{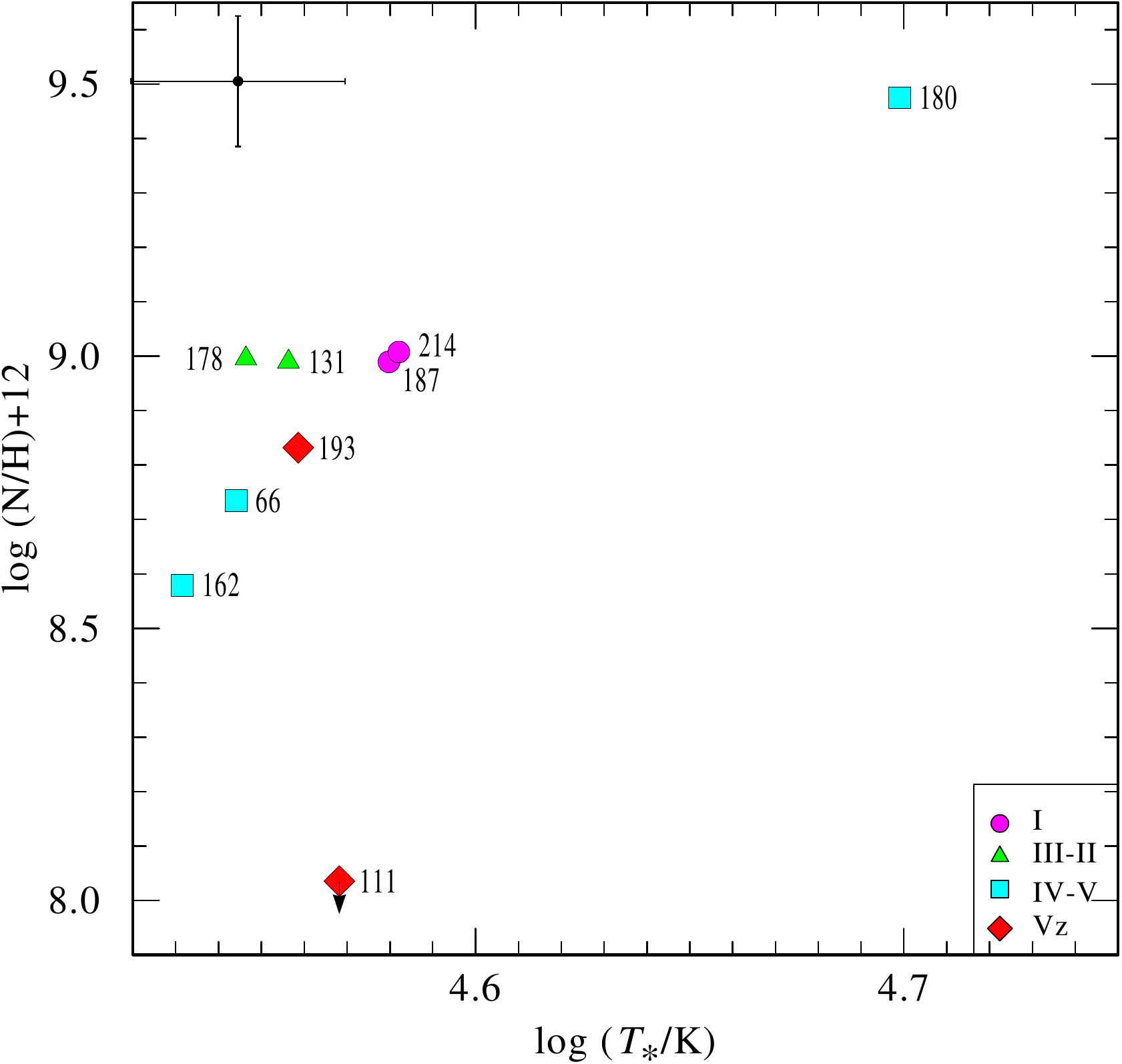}
\caption{Surface nitrogen abundances as a function of the effective temperature. 
The downward arrow indicates an upper limit to the nitrogen abundance of 
N206-FS\,111. }
\label{fig:Teff_logNH}
\end{figure}

The distribution of nitrogen abundances with effective temperature is 
illustrated in Fig\,\ref{fig:Teff_logNH}. We can see a direct correlation in our 
sample (except N206-FS\,111). In addition, all these stars show lower abundance in either 
oxygen or carbon (compared to typical LMC values) due to the CNO process. The O2 spectral type shows a very high 
nitrogen abundance in comparison to the other Of stars (but less than the total 
CNO mass fraction). In turn, its oxygen abundance measured from the \ion{O}{v} 
line at 1371\AA\, is much lower by a factor of 100 than the typical LMC values.

\begin{figure}[!ht]
\centering
\includegraphics[scale=0.48]{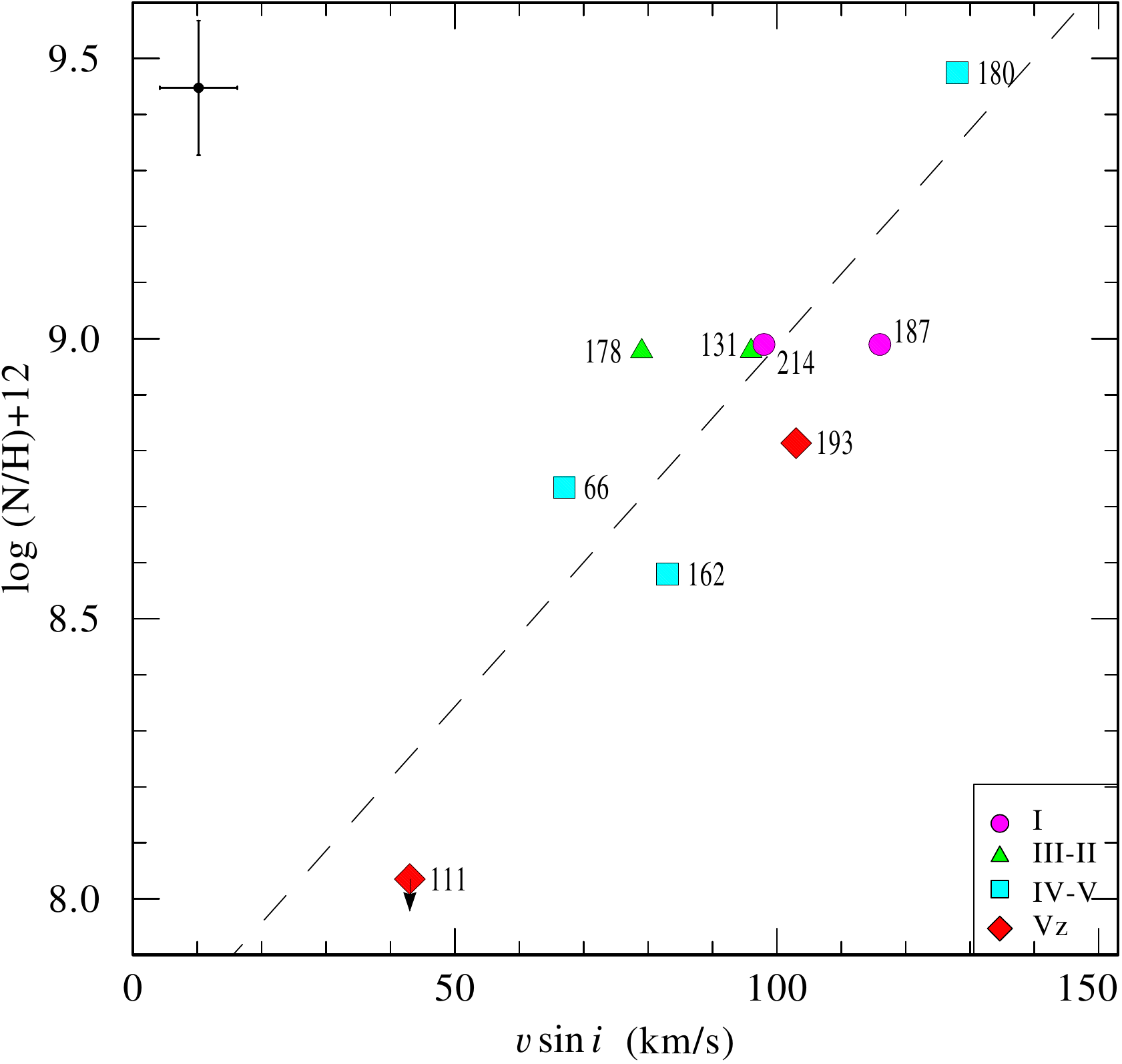}
\caption{Surface nitrogen abundances as a function of the projected rotation 
velocity $\varv$\,sin\,$i$. The dashed line is a linear regression to the data. 
The downward arrow indicates an upper limit to the nitrogen abundance of 
N206-FS\,111.}
\label{fig:vrot_logNH}
\end{figure}


One of the key questions of massive star evolution is rotational mixing and its 
impact on the nitrogen abundance. Evolutionary models accounting for rotation 
\citep{{Hunter2008},{Brott2011},{Gonzalez2012a},{Gonzalez2012b}}  predict that 
the faster a star rotates, the more mixing will occur, and the larger the 
nitrogen surface abundance that should be observed. Figure 
\,\ref{fig:vrot_logNH} shows a so-called ``Hunter-diagram'', depicting the 
variation of the nitrogen enrichment  with the projected rotational velocity. 
Our Of sample shows projected rotational velocities in the range of $\sim$40--130 
 km\,s$^{-1} $ and nitrogen abundances (log (N/H) +12) from 8 to 9.6. This 
diagram shows a relative increase in nitrogen abundance with rotation. It should 
be noted that the projected rotational velocity and the true rotational 
velocity differ by a factor sin\,$i$.
However, \citet{Maeder2009} suggests that chemical enrichment is not only a 
function of projected rotational velocity, but also depends on ages, masses, and 
metallicities. The nitrogen enriched stars are very massive in our case.

The following conclusions can be made from the diagram:
\begin{itemize}
\item The O2 dwarf N206-FS\,180 shows a exceptionally high nitrogen abundance, and 
its projected rotational velocity is highest among the sample. 
\item Evolved stars (giants and supergiants) are more nitrogen enriched than 
dwarfs of the same effective temperature.
\item The slowest rotator in our sample (also least massive), N206-FS\,111 
($\varv$\,sin\,$i$ = 43 km\,s$^{-1}$), shows very little enrichment. Although 
N206-FS\,111 and N206-FS\,193 are ZAMS candidates (Vz), N206-FS\,193 
($\varv$\,sin\,$i$ = 103 km\,s$^{-1}$) shows nitrogen enhancement.
\item The two binary candidates (N206-FS\,180a and N206-FS\,131a) are more 
nitrogen enriched than other stars with similar rotational velocities.  
\end{itemize}
\changed{
According to Kohler et al. (2015), a strong nitrogen surface enhancement can be obtained even in models with initial rotational velocities well below the threshold value. All our sample stars are nitrogen enriched with rotational velocities below 200 km/s. Because of the strong mass loss and large convective core masses, their evolutionary models above 100\,M$_{\odot}  $ shows nitrogen-enrichment irrespective of their rotation rate. The large nitrogen enhancement in our sample star N206-FS 180 is very similar(M$\approx$100\,M$_{\odot}$, $\varv$\,sin\,i$\approx$150 km/s, age$\approx$1Myr) to their model predictions. Moreover, this source is a part of binary system, which may also have influence in the enrichment.}

\subsection{Stellar feedback}
\begin{figure}[!ht]
\centering
\includegraphics[scale=0.48]{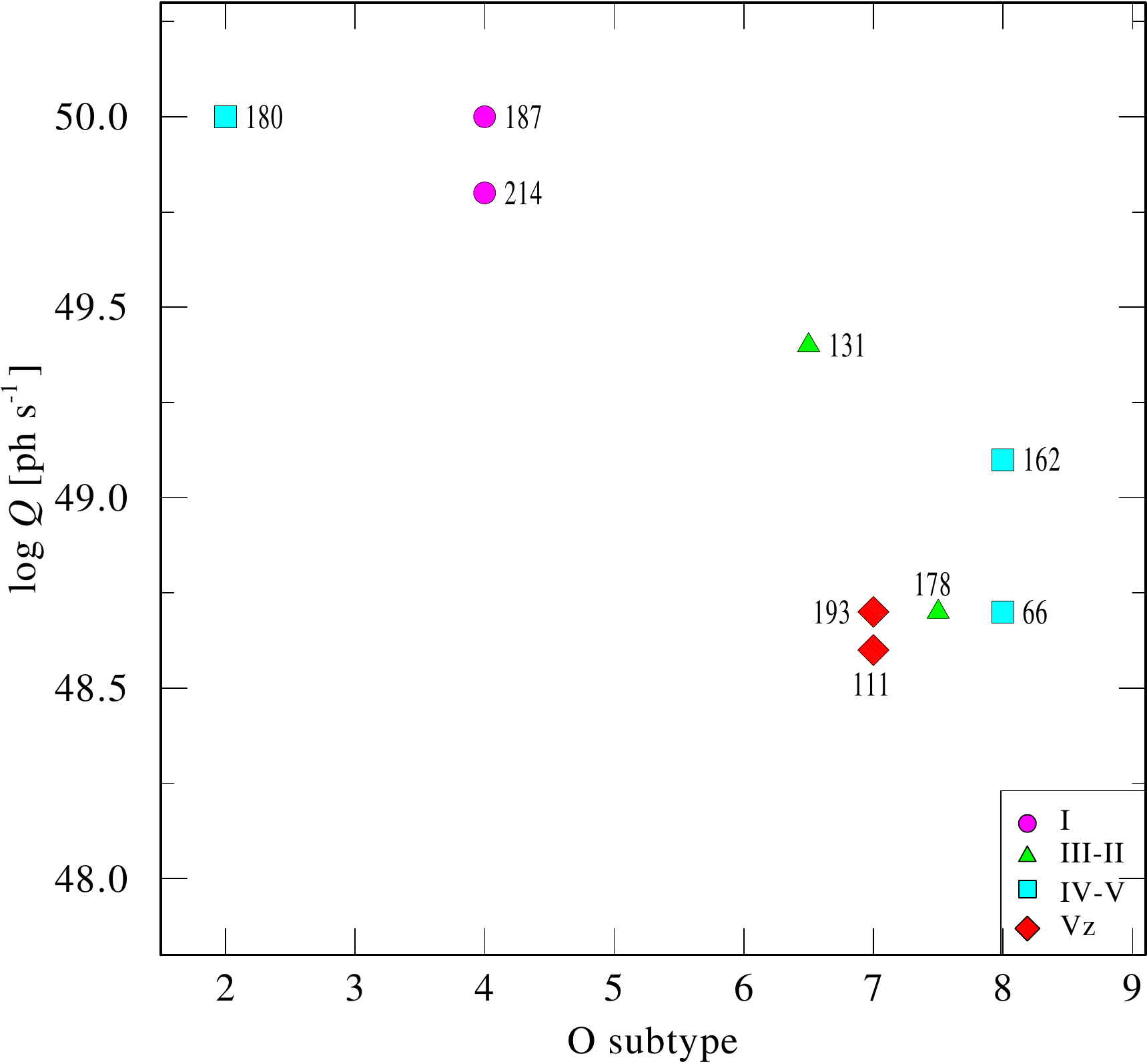}
\caption{Rate of ionizing photons as a function of the O subtype. }
\label{fig:spec_logQ}
\end{figure}


\begin{figure}[!ht]
\centering
\includegraphics[scale=0.48]{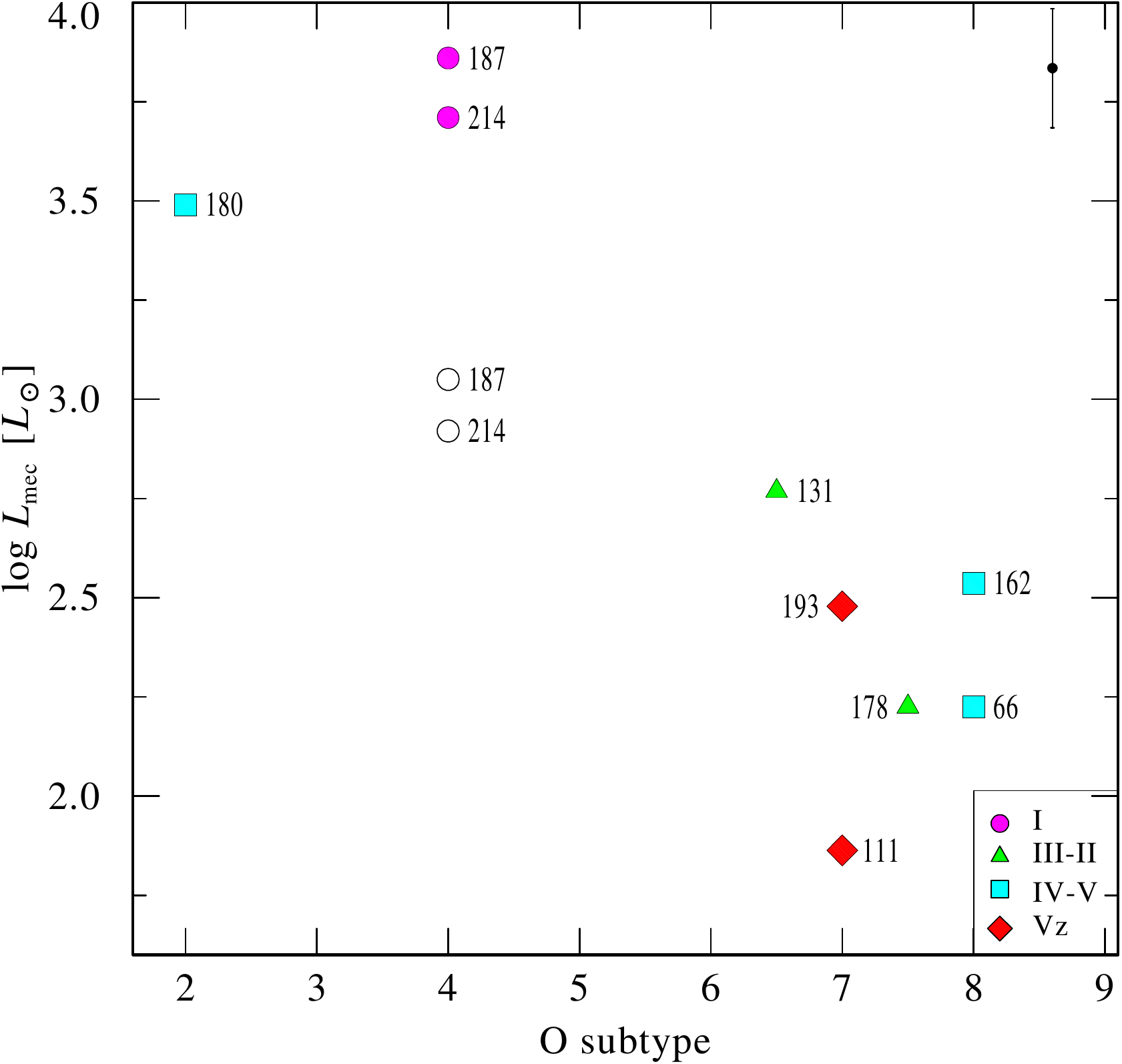}
\caption{ Mechanical luminosity of stellar winds as a function of the O 
subtype. 
 The open circles for the two supergiants refers to the models with alternative 
solution with higher clumping. Typical uncertainties are indicated with error 
bar in the upper right corner.}
\label{fig:spec_logEkin}
\end{figure}


The main contribution to the stellar feedback in the N\,206 superbubble is 
expected to come from stars in our Of sample.
The number of ionizing photons and the mechanical luminosity of the stellar 
winds as a function of O subtypes are shown in Figs.\,\ref{fig:spec_logQ} 
and \ref{fig:spec_logEkin}. Both of these energy feedback mechanisms show a 
correlation with spectral subtype, where the O2((f*)) star and the two 
supergiants dominate. The combined ionizing photon flux from the nine Of stars is 
approximately $ Q_{0}\approx\,3\times10^{50}$\,s$^{-1} $.
The total mechanical feedback from these Of stars is $ L_{\rm 
mec} = 0.5 \dot{M} \varv_\infty^{2} \approx\,1.7\times10^{4}\,L_{\odot} \,(\equiv 6.5 \times 10^{37} \rm{erg\, s^{-1}})$. We stress that the $\varv_\infty  $  was adopted for majority of the sample. The uncertainties in these estimates are propagated from mass-loss rate and terminal velocity. As now discussed, the total mechanical luminosity is dominated by the three Of stars, for which an uncertainty of approximately 30\% is estimated. Assuming stronger clumping in models would lower this total feedback by a factor of two.

\citet{Kavanagh2012} analyzed the X-ray superbubble in N\,206 using 
XMM-\textit{Newton} data. The thermal energy stored in the superbubble was estimated 
using the X-ray emitting gas, the kinetic energy of the H$\alpha$ shells, and 
the kinetic energy of the surrounding \ion{H}{i} gas. Using spectral fits 
\citet{Kavanagh2012} derived the physical properties of the hot gas and hence 
estimated the thermal energy content of the X-ray superbubble. Along with this, they 
calculated the kinetic energy of the expansion of the shell in the surrounding 
neutral gas using MCELS and 21 cm line emission line data from the ATCA-Parkes 
survey of the Magellanic Clouds \citep{Kim1998}. They estimated the total energy to be $(4.7 \pm 
1.3)\,\times 10^{51}$\,erg.

This thermal energy could be supplied  by the stars in the N206 superbubble via stellar winds and the 
supernova occurred in this region (SNR B0532-71.0). We calculated the current
mechanical energy supply from these Of stars. In order to provide the total energy estimate of these Of-type stars, we  must derive their current ages. So, we adopted the ages of individual stars from the HRD (see Sect.\,\ref{sect:hrd}) using isochrone fitting. The total mechanical energy supplied by these stars throughout their lifetimes is $(5.8 \pm 1.8)\,\times\,10^{51}$\,erg. 
As in \citet{Kavanagh2012}, we also assumed that these stars provide same amount of mechanical energy throughout their lifetime, since they are very young. Also, the time duration is short for a significant contribution from radiative and other cooling processes, and can be neglected for this rough estimate.
The energy estimates are consistent with \citet{Kavanagh2012}, even though we 
consider only the mechanical feedback from stellar winds. The total momentum 
provided from these stars over time is $\approx 1.86 \times 10^{43} {\rm 
g\,cm\,s^{-1}}$.
 
 This shows that the Of stars in our sample provide a significant contribution to 
the energy feedback.  The total feedback in the superbubble including the entire OB 
population, Wolf-Rayet stars (BAT99 53, BAT99 49), and supernova will be discussed 
in Paper II.

\section{Summary and conclusions}
\label{sect:summary}

We have obtained and analyzed FLAMES spectra of nine massive Of-type stars in 
the N\,206 giant H\,{\sc ii} complex located in the LMC. On the 
basis of these data complemented by archival UV spectra, we determined precise 
spectral types and performed a thorough spectral analysis. We used 
sophisticated PoWR model atmospheres to derive the physical and wind parameters 
of the individual stars. Our main conclusions are as follows.

\begin{itemize}

\item All stars have an age less than 4 Myr. The earliest O2 star appears to be 
less than 1\,Myr old, while the two supergiants are 2\,Myr old.
\item \changed{The wind-momentum luminosity relation of our sample are comparable to that of theoretically predicted, with the assumption of a depth dependent clumping in all models.} 
\item The star N206-FS\,187  shows very high mass-loss rate and luminosity. Its large X-ray luminosity suggests that this object is one of the X-ray brightest colliding wind binaries consisting of stars with similar spectral types.
\item Most of the Of stars are nitrogen enriched. A clear trend toward more 
enrichment with rotation as well as with temperature is found. The binary stars 
and more evolved stars show nitrogen enrichment compared to others.
\item The O2 star N206-FS\,180 shows a very high nitrogen mass fraction, while 
oxygen is strongly depleted compared to the typical LMC values.
\item The evolutionary masses are systematically lower than the corresponding 
spectroscopic masses obtained from our analysis.
\item The total ionizing photon flux and stellar wind mechanical luminosity of 
the Of stars are estimated to $3 \times 10^{50}$\,s$^{-1} $ and $1.7 \times 
10^{4}\,L_{\odot}$, respectively. This leads us to conclude that the program 
stars play a significant role in the evolution of the N\,206 superbubble.
\end{itemize}

The subsequent paper (Ramachandran et al. in prep.) will provide the detailed analysis and discussion of complete OB star population in the N\,206 superbubble.

\begin{acknowledgements}
V.R. is grateful for financial support from Deutscher Akademischer Austauschdienst (DAAD), as a part of Graduate School Scholarship Program. LMO acknowledges support by the DLR grant 50 OR 1508. A.S. is supported by the Deutsche Forschungsgemeinschaft (DFG) under grant HA 1455/26. T.S. acknowledges support from the German ``Verbundforschung'' (DLR) grant 50 OR 1612. We thank C. J. Evans for helpful discussions.
This research made use of the VizieR catalog access tool, CDS,  
Strasbourg, France. The original description of the VizieR service was  
published in A\&AS 143, 23. Some data presented in this paper were obtained from 
the Mikulski Archive for Space Telescopes (MAST). STScI is operated by the 
Association of Universities for Research in Astronomy, Inc., under NASA contract 
NAS5-26555. Support for MAST for non-HST data is provided by the NASA Office of 
Space Science via grant NNX09AF08G and by other grants and contracts.
\end{acknowledgements}

\bibliographystyle{aa}
\bibliography{paper}


\newpage

\appendix

\section{Comments on the individual stars}
\label{sect:appendixa}

\textbf{N206-FS\,66}: This O8 subgiant is located at the periphery of the 
superbubble N\,206, far away from the young cluster NGC 2018. The noticeable 
spectral features are the \ion{N}{iii} emission line (weak), the \ion{C}{iii} \& 
\ion{O}{ii} absorption blend  near 4560\,\AA, and strong \ion{Si}{iv} absorption 
lines. The model with $T _\ast$ = 36\,kK is fitted to this spectrum based on the 
\ion{He}{i}/\ion{He}{ii} line ratios. The nitrogen abundance is mainly 
constrained through the \ion{N}{iii} absorption lines in the range 
4510$-$4525\,\AA. Since the H$\alpha$ line is fully in absorption and no UV 
spectrum is available, we estimated the wind parameters with large uncertainties. 
\medskip

\textbf{N206-FS\,111}: This is one of the ZAMS type Of stars in the sample with 
spectral type O7 V((f))z. The most prominent feature in the spectrum is a very 
strong \ion{He}{ii} absorption line at 4686\,\AA. The model with $T  _\ast$ = 37 
kK is fitted to the spectrum based on the \ion{He}{i}/\ion{He}{ii} line ratios. 
This star is also described by \citet{Kavanagh2012} using the Magellanic Cloud 
Emission Line Survey (MCELS) observations. They determined its spectral type as 
O7 V  and derived a similar mass-loss rate (using theoretical relations) as in 
our analysis. The terminal velocity and the mass-loss rate have large errors, 
since H$\alpha$ is in absorption. This is the slowest rotating star in the 
sample. The nitrogen lines are very weak, hence the determined abundance is an 
upper limit.
\medskip

\textbf{N206-FS\,131}: This is one of the binary candidates in the sample. The 
presence of weak \ion{He}{i} absorption lines in the spectrum gives a clue about 
the secondary component. The strong \ion{N}{iii} emission lines and the very 
weak \ion{He}{ii}\,4686 confirm that the primary is an Of giant. The temperature of 
the primary Of is determined to be $T  _\ast$ = 38\,kK. The secondary 
component is an O star with late subtype and best fitted with a model of $T  
_\ast$ = 30\,kK. This star was mentioned as a single O8\,II star in 
\citet{Bonanos2009}. The H$\alpha$ absorption in the spectrum is partially 
filled with wind emission and is best fitted with a mass-loss rate of 3.1 $ 
\times 10^{-6} M_{\odot}\,$yr$^{-1}$. The radial velocity of this star is 
measured to be 185 km\,s$^{-1}$, which is significantly less than the systematic 
velocity of the LMC ($\approx 260$ km\,s$^{-1}$).
\medskip

\textbf{N206-FS\,162}: This object was identified as an emission line star in 
previous papers \citep{Howarth2013, Lindsay1963}. The weak \ion{N}{iii} emission 
lines in the spectrum confirm its `Of nature'. Based on the broad and very 
strong H$\alpha$ emission, we conclude that it is the only Oef-type star in the 
N\,206 superbubble. Based on the \ion{He}{i}/\ion{He}{ii} line ratios, we 
estimate a stellar temperature of $T  _\ast$ = 34\,kK. The spectrum also shows 
prominent \ion{Si}{iv} and \ion{O}{ii} absorption lines, which are used to 
constrain the abundance. Although we tried to fit the wings of  H$\alpha$ 
emission (also nitrogen emission lines), the estimated wind parameters are 
highly uncertain due to the superimposed disk emission. The projected rotational 
velocity of this star ($\varv\,\sin i \sim 83$ km\,s$^{-1}$) seems to be 
exceptionally low compared to typical Oe/Be stars. However, the morphology of the H$\alpha$ line
also suggests a nearly ``pole-on'' observer aspect.
\medskip

\textbf{N206-FS\,178}: This object is one of the giants in the sample. The spectrum is 
best fitted with a model with $T  _\ast$ = 35\,kK and log\,$g_\ast$ = 4.0. We 
mainly used the \ion{He}{i}/\ion{He}{ii} line ratio and the strong \ion{N}{iii} 
emission lines for determining the stellar temperature. Since H$\alpha$ is 
in absorption, we tried to measure the wind parameters by adjusting the  nitrogen 
emission line fits. For this star, we varied the abundance N in the model to best fit the \ion{N}{iii} absorption lines.
\medskip

\textbf{N206-FS\,193}:  This is one of the stars located at the center of the 
cluster NGC 2018. This star is also tabulated in \citet{Kavanagh2012}, but with 
spectral type O8 V. We classified this star as O7 Vz. The spectrum with weak 
\ion{N}{iii} emission lines and a very strong \ion{He}{ii}\,4686  absorption 
line is best fitted with a model that has a temperature of $T  _\ast$ 
= 36\,kK. This spectrum is highly contaminated with nebular emission.
\medskip

\textbf{N206-FS\,214}: An Of supergiant in the sample. \citet{Massey1995} 
catalogued this star as an O4 If spectral type. \citet{Graczyk2011} identified 
this object as an eclipsing binary (OGLE-LMC-ECL-18366) with a period of 64.9 days 
based on the data from Optical Gravitational Lensing Experiment (OGLE). Their light curve suggests 
that both components have similar temperatures (from the depth) and the primary has a
much larger radius (from the width). We identified this spectrum as an O4 If supergiant, with 
no trace of a secondary component in the optical spectrum. This could be an indication of a very faint secondary 
(dwarf/giant). 

\begin{figure*}
\centering
\includegraphics[scale=0.8]{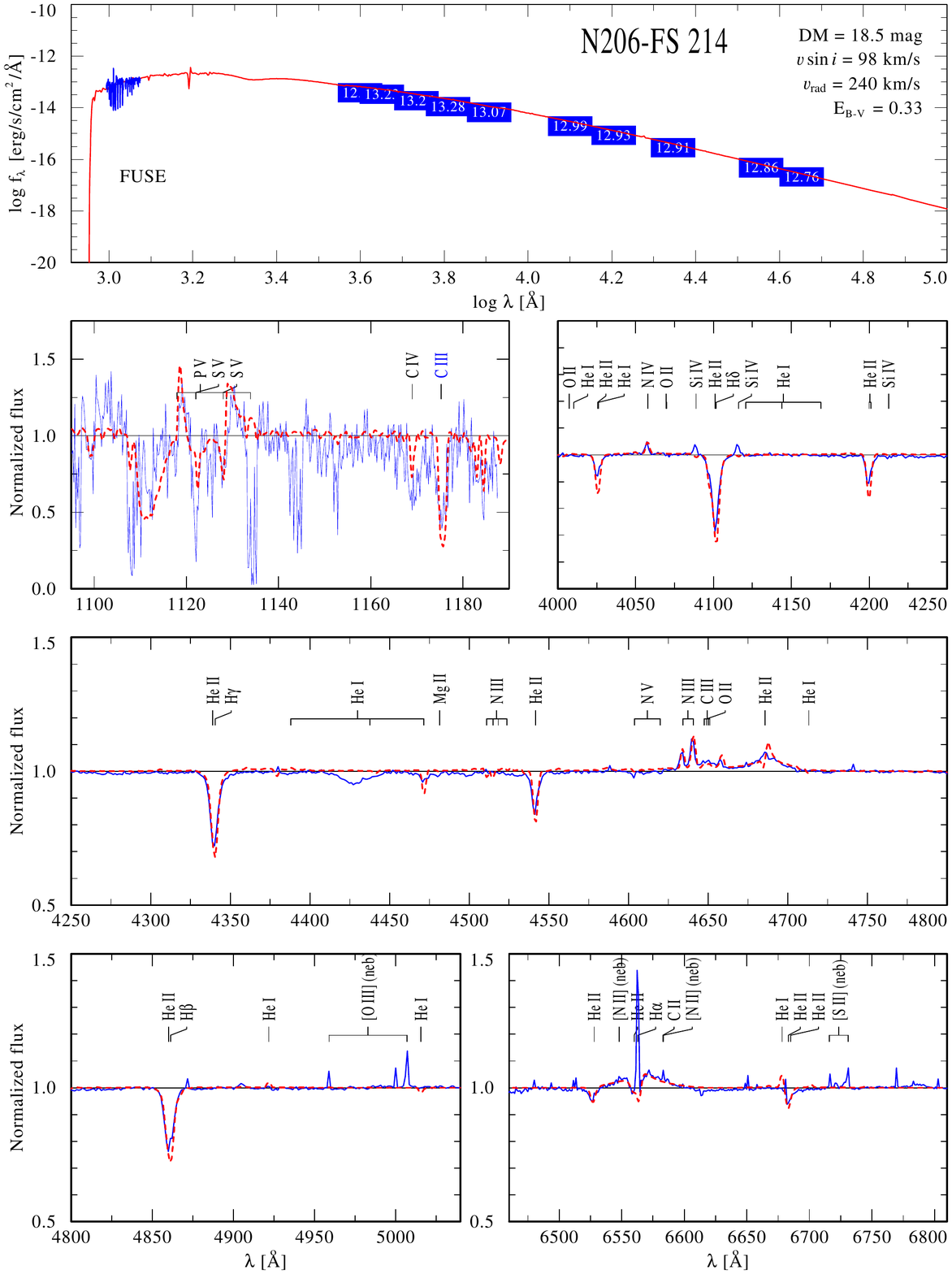}
\caption{Spectral fit for N206-FS\,214. The SED is best fitted to flux 
calibrated FUSE FUV spectra. The second panel shows the FUSE spectrum normalized 
with the model continuum. }
\label{fig:RHH214}
\end{figure*}

The spectral features of this object are similar to N206-FS\,187, and both are 
best fitted with the same model. A FUSE spectrum is available for this object, 
which is shown in the second panel of Fig.\,\ref{fig:RHH214}. The prominent UV 
P-Cygni profiles \ion{P}{v\,$\lambda\lambda$1118--1128} and 
\ion{S}{v\,$\lambda\lambda$1122--1134} are used to measure the wind parameters 
along with H$\alpha$. In order to best fit the \ion{P}{v} lines, we used a model 
with stronger clumping ($D=20$ and $R_{\rm D} = 0.05\,R_\ast$) and lower 
mass-loss rate, i.e., a factor of 4 lower than a model with default clumping, similar 
to N206-FS\,187. The carbon lines \ion{C}{iv} at 1169\,\AA\, and \ion{C}{iii} at 
1176\,\AA\ show strong absorption. The calibrated UV spectrum is normalized and 
fitted consistently with the model SED, and hence constrained the luminosity and 
reddening values of this star at high precision.

\section{Spectral fitting}
\label{sect:appendixb}
In this section, we present the spectral fits of all stars analyzed
in this study. The upper panel shows the SED with 
photometry from UV, optical and infrared bands. Lower panels show the normalized VLT-FLAMES spectrum 
depicted by blue solid lines. The observed spectrum is overplotted with PoWR 
spectra shown by red dashed lines. Main parameters of the best-fit model are 
given in Table\,\ref{table:stellarparameters}.

 \begin{figure*}
\centering
\includegraphics[scale=0.8]{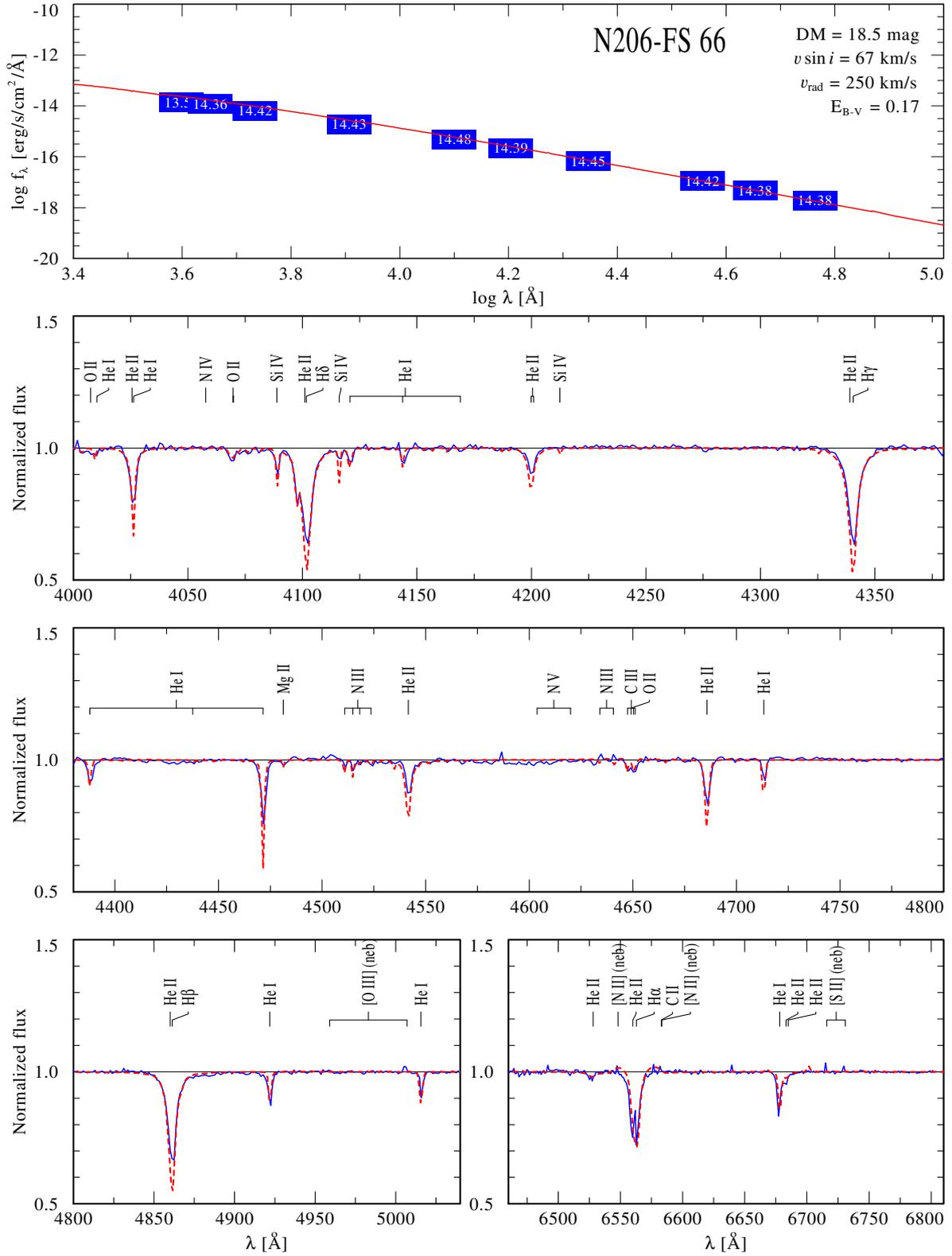}
\caption{Spectral fit for N206-FS\,66}
\label{fig:RHH66}
\end{figure*}

 \begin{figure*}
\centering
\includegraphics[scale=0.8]{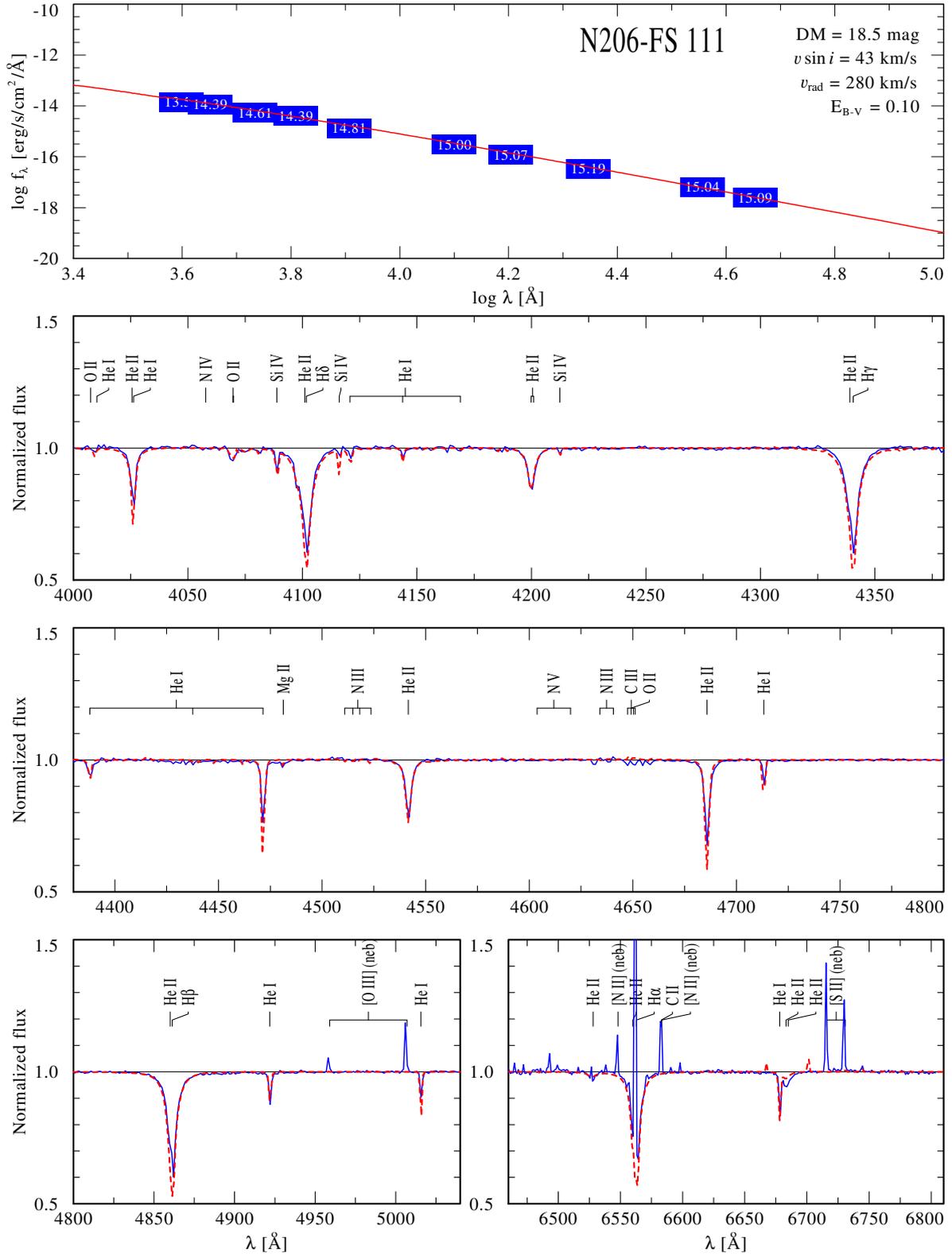}
\caption{Spectral fit for N206-FS\,111}
\label{fig:RHH111}
\end{figure*}

 \begin{figure*}
\centering
\includegraphics[scale=0.8]{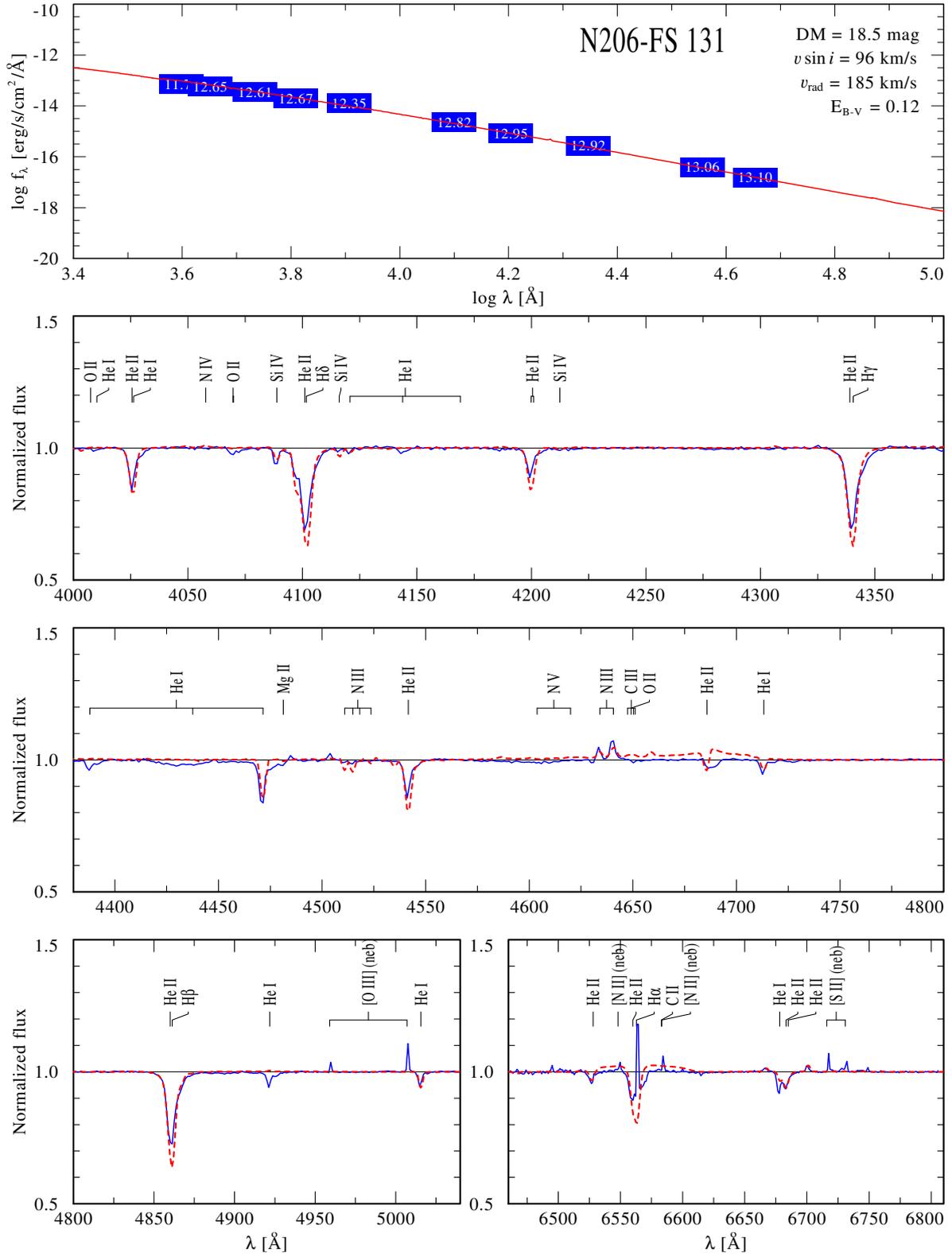}
\caption{Spectral fit for N206-FS\,131, which is suspected to be a binary (see 
Appendix A and next figure)}
\label{fig:RHH131}
\end{figure*}

 \begin{figure*}
\centering
\includegraphics[scale=0.8]{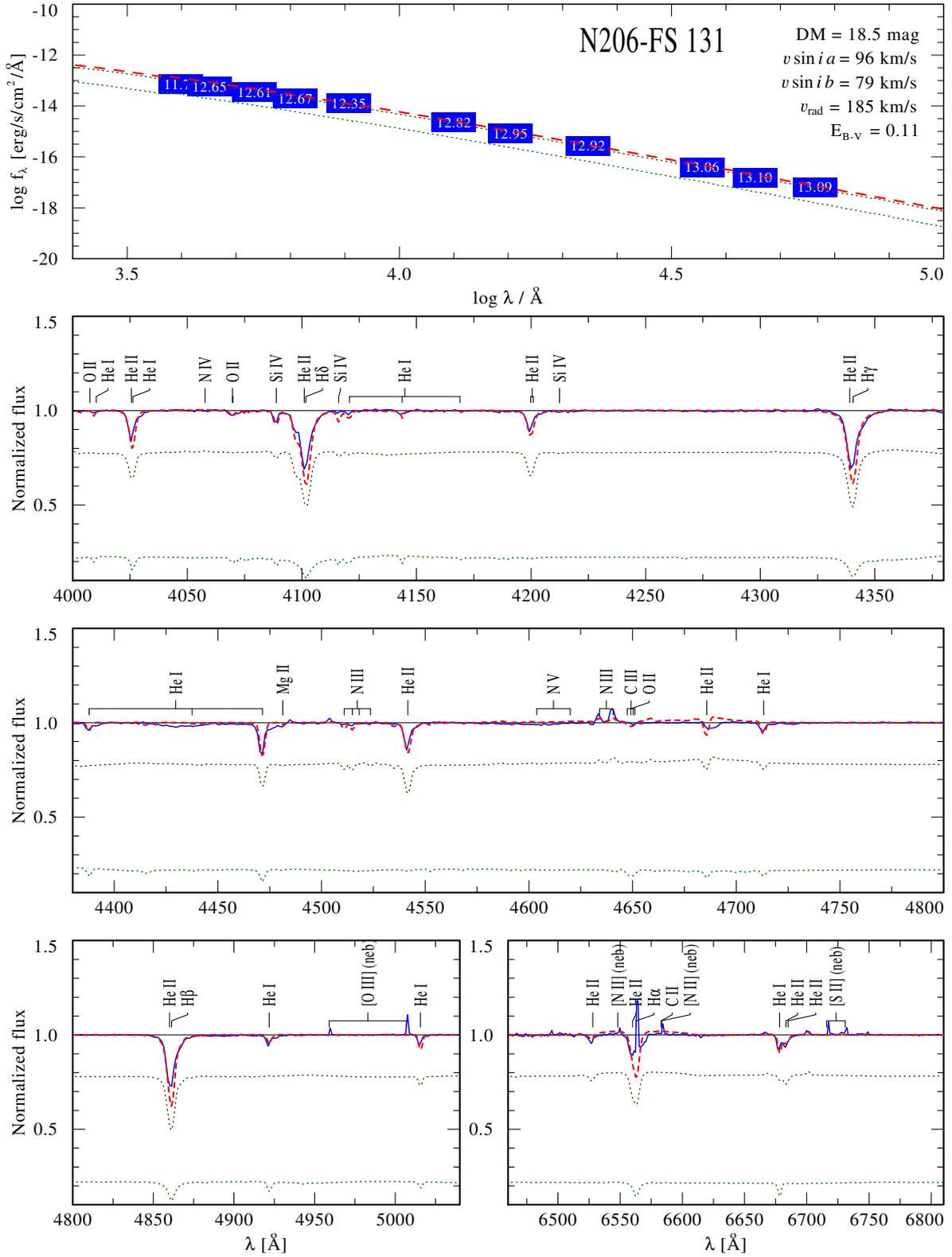}
\caption{Composite model (red dashed line) for N206-FS\,131. The primary 
Of star (brown dotted line) is fitted with $T  _\ast$ = 38 kK model and 
secondary component (green dotted line) to a model with $T  _\ast$ = 30\,kK. See 
Table\,\ref{table:stellarparameters} for more information.}
\label{fig:RHH131b}
\end{figure*}

 \begin{figure*}
\centering
\includegraphics[scale=0.8]{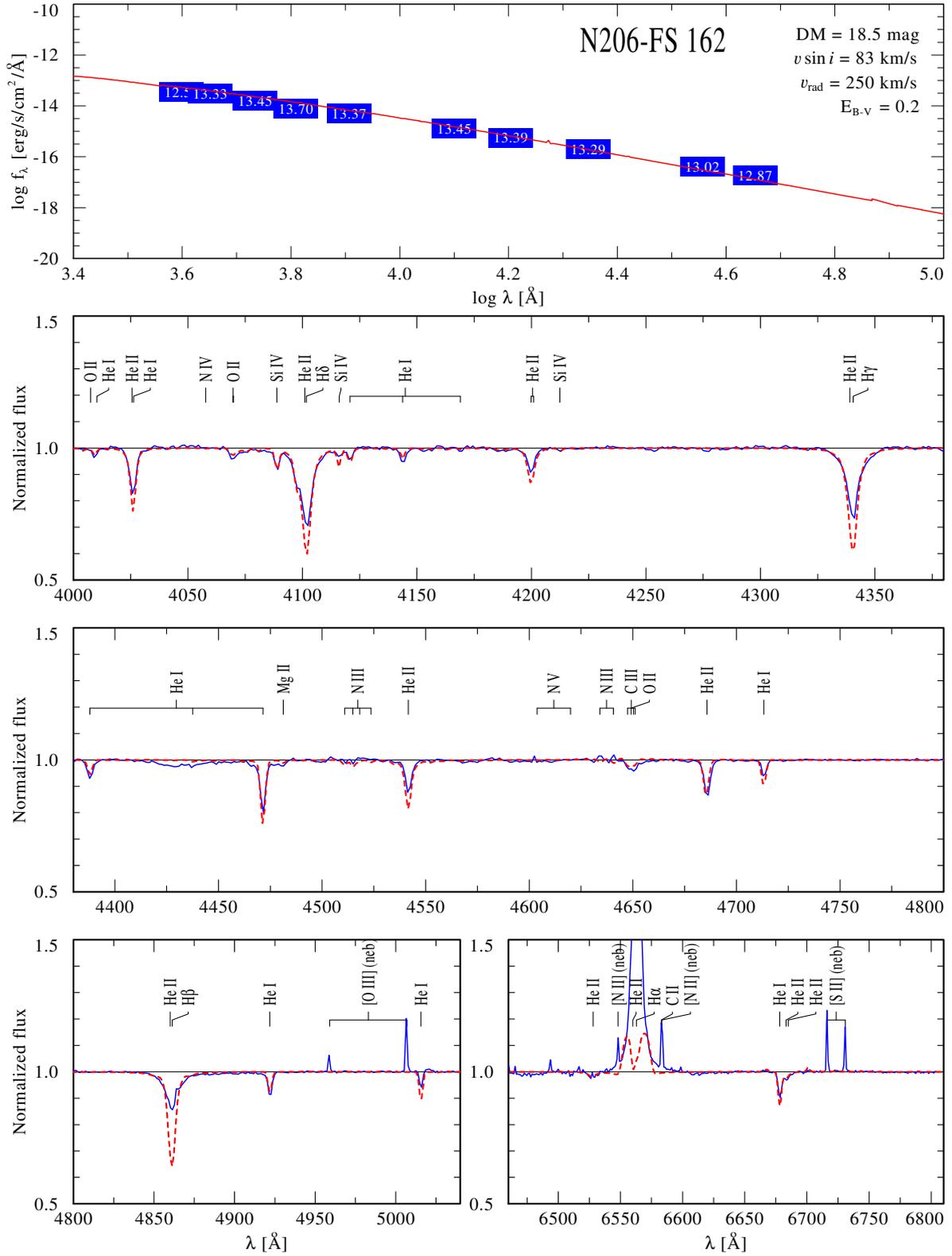}
\caption{Spectral fit for N206-FS\,162}
\label{fig:RHH162}
\end{figure*}

\begin{figure*}
\centering
\includegraphics[scale=0.8]{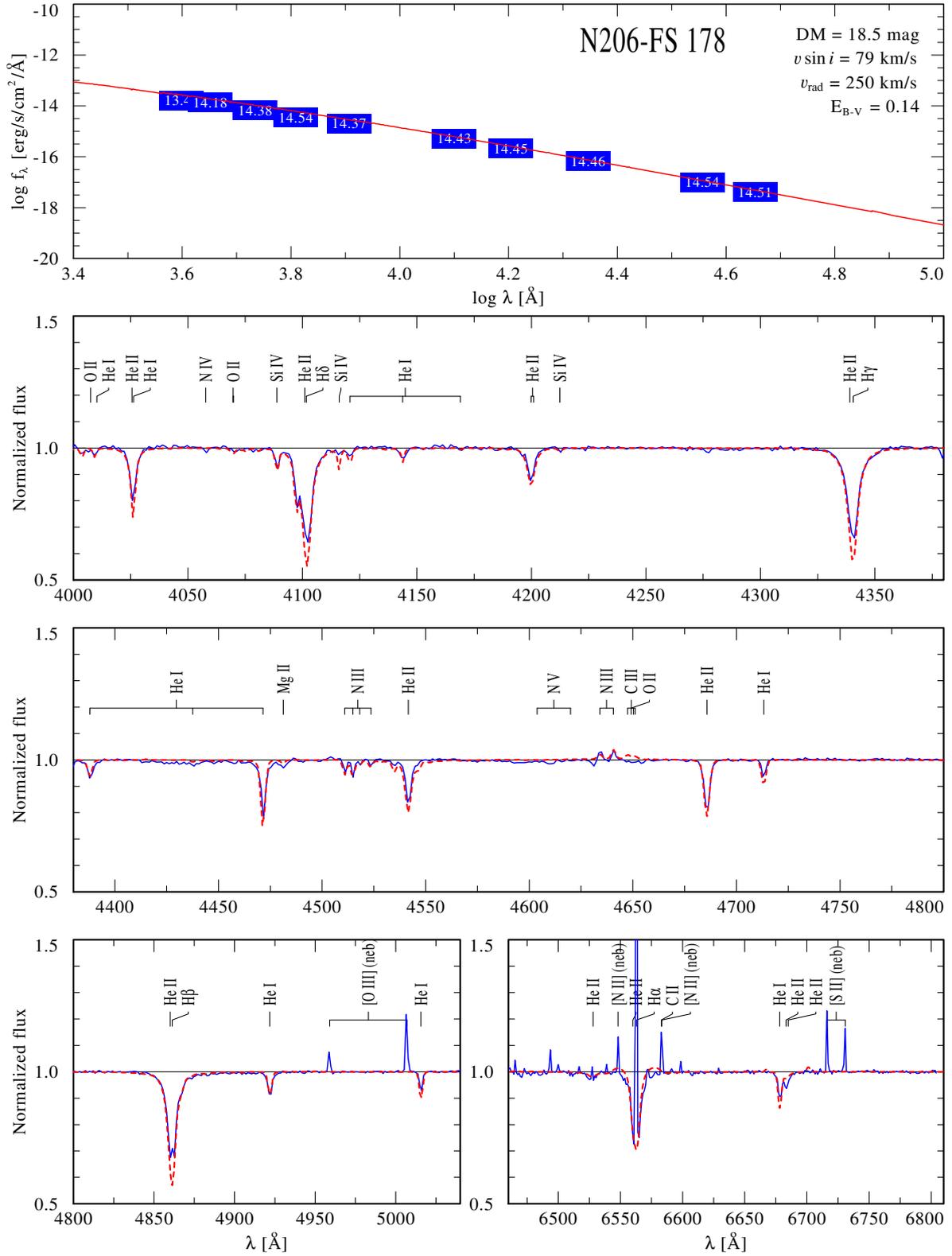}
\caption{Spectral fit for N206-FS\,178.}
\label{fig:RHH178}
\end{figure*}

\begin{figure*}
\centering
\includegraphics[scale=0.8]{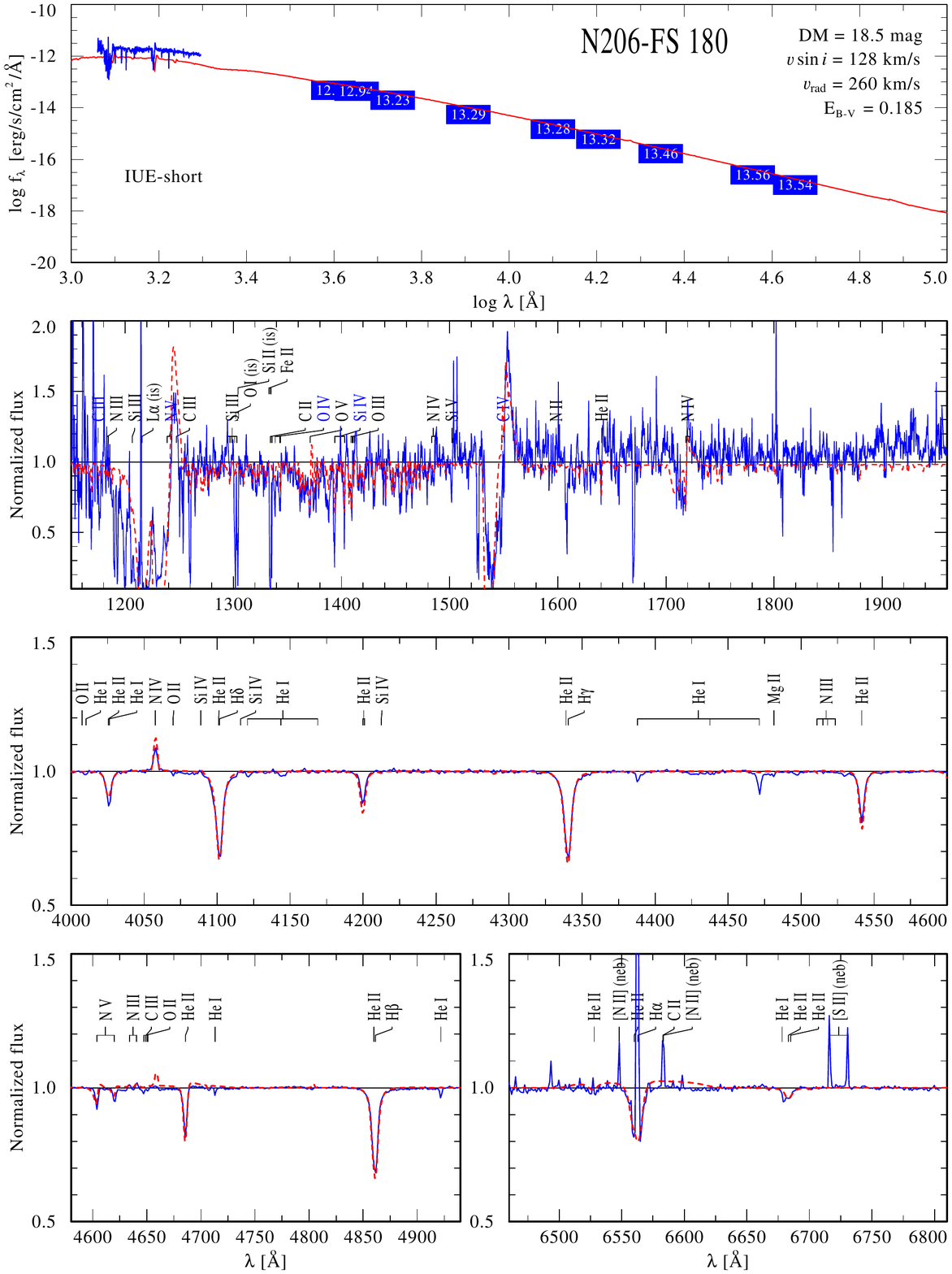}
\caption{Spectral fit for N206-FS\,180 with a single star model. The \ion{He}{i} lines in the observation are absent in the model, indicating the 
contamination from a secondary component. See next figure for the composite 
model.}
\label{fig:RHH180}
\end{figure*}

 \begin{figure*}
\centering
\includegraphics[scale=0.8]{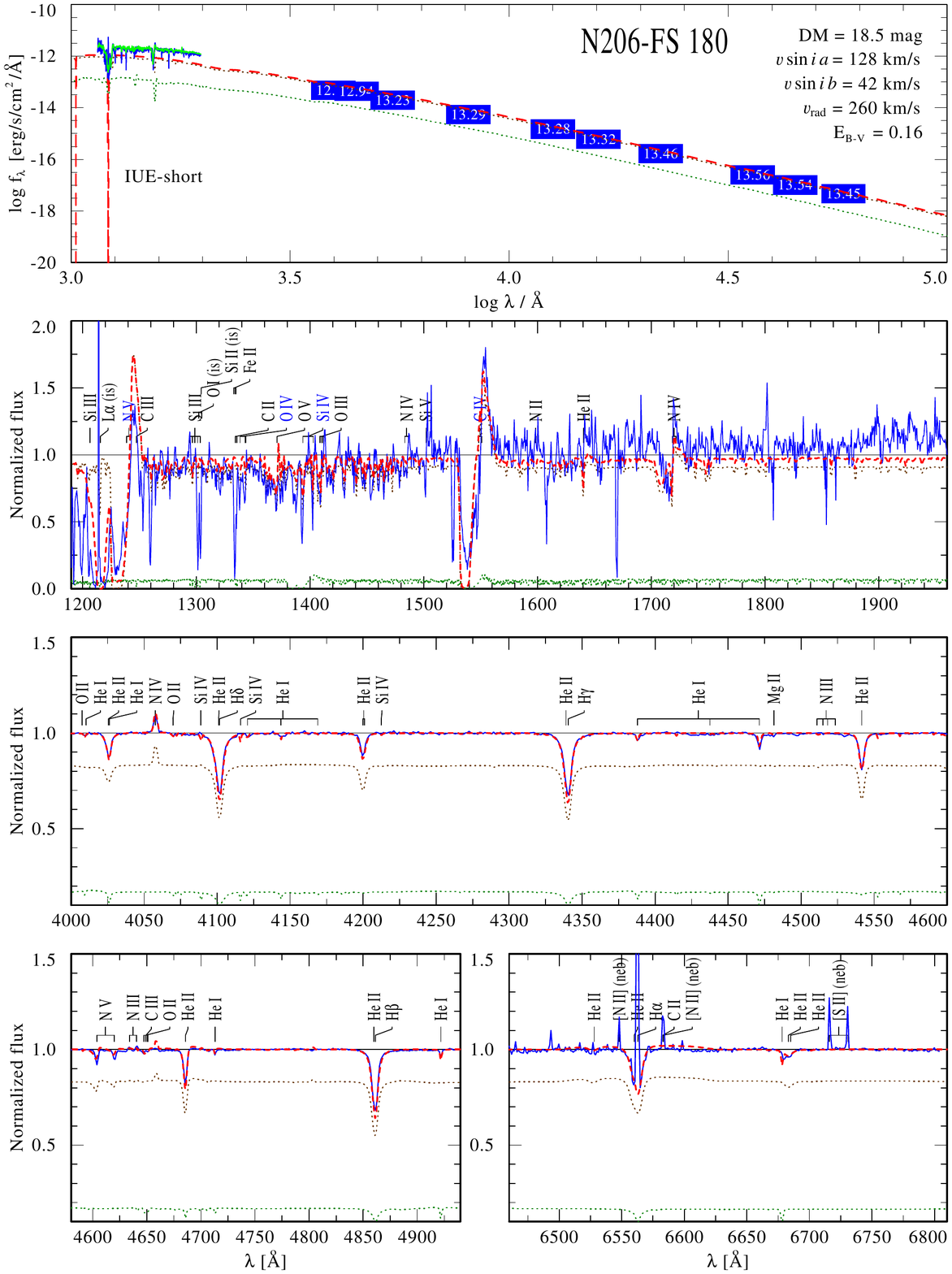}
\caption{Composite model (red dashed line) for N206-FS\,180 is the weighted 
sum of O2 V((f*)) (brown dotted line) and O8 IV (green dotted line) model 
spectra with effective temperature 50\,kK and 32\,kK, respectively. The relative 
offsets of the model continua correspond to the light ratio between the two 
stars. See Table\,\ref{table:stellarparameters} for more details.}
\label{fig:BinRHH180}
\end{figure*}

 \begin{figure*}
\centering
\includegraphics[scale=0.8]{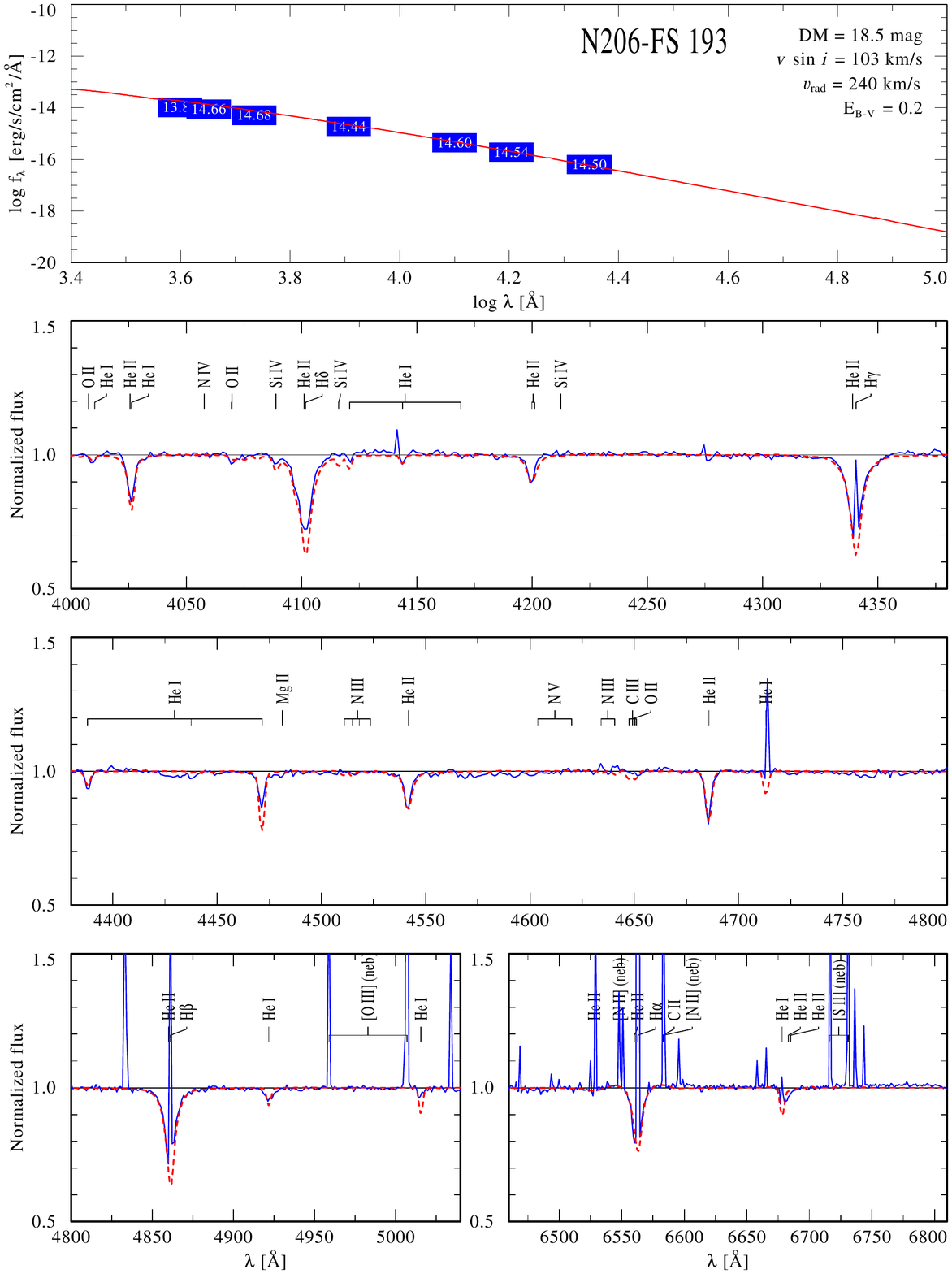}
\caption{Spectral fit for N206-FS\,193}
\label{fig:RHH193}
\end{figure*}

\end{document}